\documentclass[12pt]{iopart}

\usepackage{microtype, xspace, amssymb, bm, mathabx, tabularx,enumitem}

\renewcommand{\theenumi}{(\roman{enumi})}

\usepackage{geometry}
\usepackage{xfrac}
\usepackage[normalem]{ulem}
\usepackage[dvipsnames, usenames]{xcolor}
\definecolor{linkcolor}{rgb}{0.0,0.3,0.5}
\definecolor{dodgerblue}{HTML}{1E90FF}
\usepackage[unicode, colorlinks=true, linkcolor=linkcolor, citecolor=linkcolor, filecolor=linkcolor,urlcolor=linkcolor, pdfusetitle]{hyperref}
\usepackage[all]{hypcap}
\usepackage[T1]{fontenc}
\usepackage[utf8]{inputenc}
\usepackage{orcidlink}
\usepackage{tensor}
\usepackage{multirow}
\usepackage{rotating}
\usepackage{booktabs}
\usepackage[compress]{cite}

\newcommand{\num}[1]{#1}

\newcommand{\ett}  {$\texttt{ET-}\triangle$\xspace}
\newcommand{\ettp} {$\texttt{ET-}\triangle + \texttt{I+}$\xspace}
\newcommand{\etts} {$\texttt{ET-}\triangle + \texttt{I\#}$\xspace}
\newcommand{\ettc} {$\texttt{ET-}\triangle + \texttt{CE}$\xspace}

\newcommand{\etol}  {$\texttt{ET-L}$\xspace}
\newcommand{\etl}  {$\texttt{ET-2L}$\xspace}
\newcommand{\etlp} {$\texttt{ET-2L} + \texttt{I+}$\xspace}
\newcommand{\etls} {$\texttt{ET-2L} + \texttt{I\#}$\xspace}
\newcommand{\etlc} {$\texttt{ET-2L} + \texttt{CE}$\xspace}

\newcommand{\ce}   {$\texttt{CE}$\xspace}
\newcommand{\cep}  {$\texttt{CE} + \texttt{I+}$\xspace}
\newcommand{\ces}  {$\texttt{CE} + \texttt{I\#}$\xspace}
\newcommand{\cet}  {$\texttt{CE} + \texttt{ET-}\triangle$\xspace}
\newcommand{\cetp} {$\texttt{CE} + \texttt{ET-}\triangle + \texttt{I+}$\xspace}
\newcommand{\cets} {$\texttt{CE} + \texttt{ET-}\triangle + \texttt{I\#}$\xspace}
\newcommand{\cel}  {$\texttt{CE} + \texttt{ET-2L}$\xspace}
\newcommand{\celp} {$\texttt{CE} + \texttt{ET-2L} + \texttt{I+}$\xspace}
\newcommand{\cels} {$\texttt{CE} + \texttt{ET-2L} + \texttt{I\#}$\xspace}

\newcommand{\indp}  {$\texttt{I+}$\xspace}
\newcommand{\inds}  {$\texttt{I\#}$\xspace}

\newcommand{\tobs}    {\ensuremath{T_{{\rm obs}}}\xspace}
\newcommand{\nth}     {\ensuremath{N_{\rm th}}\xspace}
\newcommand{\snrth}   {\ensuremath{\rho_{\rm th}}\xspace}
\newcommand{\skyth}   {\ensuremath{\Omega_{\rm th}}\xspace}
\newcommand{\disth}   {\ensuremath{(\Delta D_{L} / D_L)_{\rm th}}\xspace}
\newcommand{\volth}   {\ensuremath{V_{\rm th}}\xspace}
\newcommand{\pmsnrth} {\ensuremath{\rho_{\rm pm,th}}\xspace}
\newcommand{\ewsnrth} {\ensuremath{\rho_{\rm ew,th}}\xspace}
\newcommand{\ewskyth} {\ensuremath{\Omega_{\rm ew,th}}\xspace}
\newcommand{\fpm}     {\ensuremath{f_{\rm pm}}\xspace}
\newcommand{\zloth}   {\ensuremath{\Delta z_{\rm\vee,th}}\xspace}
\newcommand{\PIcurve}   {\ensuremath{\mathcal{P}_{\mathcal{I}}}\xspace}

\newcommand{\mbbh}     {^{\rm BBH}}
\newcommand{\mbns}     {^{\rm BNS}}

\newcommand{\milan}{Dipartimento di Fisica ``G. Occhialini'', Universit\'a degli Studi di Milano-Bicocca, Piazza della Scienza 3, 20126 Milano, Italy}
\newcommand{\infn}{INFN, Sezione di Milano-Bicocca, Piazza della Scienza 3, 20126 Milano, Italy}

\definecolor{excol}{HTML}{7570b3}

\begin{document}

\makeatletter
\setcounter{footnote}{0}
\renewcommand{\thefootnote}{\arabic{footnote}}
\renewcommand{\@makefnmark}{\hbox{\textsuperscript{\@thefnmark}}}
\renewcommand{\@makefntext}[1]{%
  \noindent\hbox{\textsuperscript{\@thefnmark}}#1%
}
\makeatother


\title[Impact of facility timing and coordination for XG GW detectors]{Impact of
facility timing and coordination for next-generation gravitational-wave
detectors} 
 
\author{    Ssohrab Borhanian$^{1,2}\,$\orcidlink{0000-0003-0161-6109},
            Arianna Renzini$^{1,2,3}\,$\orcidlink{0000-0002-4589-3987},
            \\Philippa S. Cole$^{1,2}\,$\orcidlink{0000-0001-6045-6358},
            Costantino Pacilio$^{1,2}\,$\orcidlink{0000-0002-8140-4992},
            \\Michele Mancarella$^{4}\,$\orcidlink{0000-0002-0675-508X}, and
            Davide Gerosa$^{1,2}\,$\orcidlink{0000-0002-0933-3579}}
\address{$^1$ \milan}
\address{$^2$ \infn}
\address{$^3$ Department of Physics, ETH Z\"urich, Wolfgang-Pauli-Strasse 27, 8093 Z\"urich, Switzerland}
\address{$^4$ Aix-Marseille Universit\'e, Universit\'e de Toulon, CNRS, CPT, Marseille, France}
\ead{ssohrab.borhanian@unimib.it}

\vspace{10pt}
\begin{indented}
\item[]\today
\end{indented}

\begin{abstract}
While the Einstein Telescope and Cosmic Explorer proposals for next-generation,
ground-based detectors promise vastly improved sensitivities to
gravitational-wave signals, only joint observations are expected to enable the
full scientific potential of these facilities, making timing and coordination
between the efforts crucial to avoid missed opportunities. 
This study investigates the impact of long-term delays on the scientific
capabilities of next-generation detector networks. We use the Fisher
information formalism to simulate the performance of a set of detector networks
for large, fiducial populations of binary black holes, binary neutron stars,
and primordial black-hole binaries. Bootstrapping the simulated populations, we
map the expected observation times required to reach a number of observations
fulfilling scientific targets for key sensitivity and localization metrics
across various network configurations. We also investigate the sensitivity to
stochastic backgrounds.
We find that purely sensitivity-driven metrics such as the signal-to-noise
ratio are not strongly affected by delays between facilities. This is
contrasted by the localization metrics, which are very sensitive to the number
of detectors in the network and, by extension, to delayed observation campaigns
for a detector.  Effectively, delays in one detector behave like network-wide
interruptions for the localization metrics for networks consisting of two
next-generation facilities.  We examine the impact of a supporting,
current-generation detector such as LIGO India operating concurrently with
next-generation facilities and find such an addition will greatly mitigate the
negative effects of delays for localization metrics, with important
consequences on multi-messenger science and stochastic searches.
\end{abstract}

%

\section{Introduction}

Gravitational-wave (GW) astronomy currently relies on four ground-based detector
facilities: the LIGO detectors~\cite{LIGOScientific:2014pky} in Hanford,
Washington, USA and Livingston, Louisiana, USA, the Virgo detector
\cite{TheVirgo:2014hva} in Cascina, Italy, and the KAGRA detector
\cite{Aso:2013eba} in Hida, Japan. A third LIGO
detector~\cite{Unnikrishnan:2013qwa} is under construction in Aundha,
Maharashtra, India, adding the fifth and final piece to the global network of
second-generation, ground-based detectors. The LIGO, Virgo, and KAGRA detectors
observed about 218 compact binary coalescence candidates, including binary black
holes (BBHs), binary neutron stars (BNSs), and neutron star-black hole
mergers~\cite{LIGOScientific:2018mvr, LIGOScientific:2020ibl,
LIGOScientific:2021djp, LIGOScientific:2025slb} during their first three
observing runs and the initial part of the fourth run, O4a. Additional data
releases for the subsequent parts will follow.

Amidst the excitement of the early years of GW astronomy the community has
started to plan for potential upgrades to existing facilities, such as the
proposed upgrade to the LIGO detectors from A+ to A\# sensitivity
\cite{KAGRA:2013rdx,asharp}, in addition to proposals for next-generation (XG),
ground-based GW detectors. Two proposals are currently gathering the most
attention: the Einstein Telescope (ET)~\cite{Punturo:2010zz, Abac:2025saz} and
Cosmic Explorer (CE)~\cite{Reitze:2019iox}. Both proposed XG facilities are
expected to dramatically improve the GW sensitivity across the audio band
ranging from a few to order a  thousand Hz, enabling the observation of hundreds
of thousands of BBH and BNS signals per year and at much higher fidelities
\cite{Vitale:2016icu, Singh:2021zah, Borhanian:2022czq, Iacovelli:2022bbs,
Ronchini:2022gwk, Banerjee:2022gkv, Yi:2022tzs, Gupta:2023evt,
Santoliquido:2024oqs}. 

Both proposing teams are still narrowing down the exact configurations of the
detector facilities. There are several options under consideration for
ET~\cite{Branchesi:2023mws,Abac:2025saz} including one triangular configuration
with interferometer arm lengths of $10\,{\rm km}$ in either Sardinia, Italy, or
the Meuse-Rhine region between the Netherlands, Belgium, and Germany, as well as
two L-shaped configurations with varying arm lengths at both of the
aforementioned sites.\footnote{Recently, a third candidate site in Saxony
(Germany) has also been under active consideration, but it is not further
investigated in this study, in line with Ref.~\cite{Branchesi:2023mws}.}
Similarly, a range of CE~\cite{Evans:2021gyd, Evans:2023euw} configurations are
being explored which are centered around one or two L-shaped interferometers
with arm lengths of $20\text{ -- }40\,{\rm km}$ that could be built in the
United States and/or Australia.

Next-generation GW detectors require substantial investment and would stand
among the foremost frontier science facilities of the coming decades.
Therefore, coordination between efforts such as ET and CE will be crucial to
maximize their scientific return. Conversely, non-optimal synchronization could
represent a major missed opportunity for both astrophysics and fundamental
physics. This study aims to investigate the impact of facility timing---such as
planning and construction delays---on the scientific capabilities of XG
detector networks, focusing on the expected observation times for key metrics
in GW astronomy. In particular, some science targets can be achieved relatively
quickly by a single XG facility, potentially affecting the scientific rationale
for the other facility if it is delayed. Conversely, while other science goals
can be reached orders of magnitude faster only if both the ET and CE facilities
operate concurrently. In the latter case, the still-under-construction
LIGO-India detector can play a crucial role to mitigate the studied delays when
operating in tandem with future XG
networks~\cite{Vitale:2018nif,Pandey:2024mlo}.

Section~\ref{sec:methods} describes the science metrics used in this study, the
XG networks investigated, and the specifics of the simulated sources.
Section~\ref{sec:results} presents the impact of XG facility timing on their
scientific return, the impact of LIGO-India, the discernability of primordial
origin, and the sensitivity to stochastic backgrounds. Finally,
Sec.~\ref{sec:conclusion} summarizes and discusses the findings, implications,
and caveats of this study.

\section{Setup} \label{sec:methods}

\subsection{Targeted metrics} 

This study aims to map expectations for observation times required to achieve
targeted thresholds for key science metrics in GW astronomy such as
signal-to-noise ratio (SNR) and the estimation fidelity in luminosity distance,
sky position, and 3D localization; particularly taking into account that
detector facilities in networks could commence observation with {\em delays}
with respect to each other. A variety of XG detector networks are examined, as
well as \num{three} types of compact-binary GW sources, namely BBHs and BNSs of
stellar origin and primordial black hole (PBH) binaries of cosmological origin.

The quantities of interest are the expected observation times \tobs to reach a
number of events \nth satisfying the targeted thresholds for \num{nine} key
science metrics: 
\begin{enumerate}[leftmargin=1.2cm]
\item full-signal SNR \snrth, 
\item post-merger SNR \pmsnrth (for BBHs),
\item early-warning SNR \ewsnrth (for BNSs),
\item 90\%-credible sky area \skyth, 
\item 90\%-credible sky area from early warnings~\ewskyth (for BNSs),
\item relative luminosity distance error \disth,
\item comoving error volume \volth, 
\item lower redshift error bound \zloth (for PBH binaries),
\item power-law sensitivity \PIcurve of a set of detectors $\mathcal{I}$ to a
stochastic signal.  \end{enumerate}
These metrics are designed to span the full breadth of GW science, from
compact-binary astrophysics to multi-messenger studies and tests of General
Relativity.  In particular, the BBH post-merger regime is defined, for the
purposes of this study, to start at a frequency \fpm corresponding to the peak
of the GW strain in the frequency domain such that
\begin{eqnarray}
\left(\left|\tilde{h}_+ - {\rm i}\,
\tilde{h}_\times\right|\,f^{7/6}\right)_{f=f_{\rm pm}} =
\max_{f}\left[\left|\tilde{h}_+ - {\rm i}\,
\tilde{h}_\times\right|\,f^{7/6}\right]\,,
\end{eqnarray}
where $h_+,h_\times$ are the Fourier transforms of the plus and cross polarizations
of the GW strain. For BNS, the early-warning time is set to \num{$\tau_{\rm
ew}=2\,{\rm min}$} before coalescence at which the signal has reached a
frequency
\begin{eqnarray}
f_{\rm ew} = 22.2 \times
\left(\frac{0.25}{\eta}\right)^{3/8}\left(\frac{2.8\,M_\odot}{M}\right)^{5/8}
\left(\frac{120\,{\rm s}}{\tau_{\rm ew}}\right)^{3/8} \,{\rm Hz},
\end{eqnarray}
with symmetric mass ratio $\eta=m_1m_2 / M^2$ and total detector-frame mass of
the system $M=m_1+m_2$.  The comoving error volume is computed from the sky
area $\Omega$ and luminosity distance error $\Delta D_L$ as
\begin{eqnarray}
V = \Omega  \int\limits_{\Delta z_\vee}^{\Delta z^\wedge} \mathcal{V}_{c}(z)\,
{\rm d}z,
\end{eqnarray}
where the differential comoving volume $\mathcal{V}_{c} \, {\rm [Mpc^3/rad^2]}$
as well as the lower and upper redshift error bounds $\Delta z_\vee=z[D_L -
\Delta D_L]$ and $\Delta z^\wedge=z[D_L + \Delta D_L]$ are computed using the
\num{\texttt{Planck18}} cosmology~\cite{Planck:2018vyg} with redshift $z[\cdot]$ as a
function of luminosity distance.

The metrics (i) - (viii) are computed for \num{$10^6$} simulated BBH, BNS, and
PBH binary signals using the open-soure Python package \texttt{gwbench}
\cite{Borhanian:2020ypi} which implements the Fisher-matrix formalism
\cite{Cutler:1994ys, Poisson:1995ef, Balasubramanian:1995bm}. The formalism is
well-established in GW astronomy for the purposes of benckmarking the science
capabilties of next-generation GW detectors for a large number of signals
\cite{Borhanian:2022czq, Iacovelli:2022bbs, Ronchini:2022gwk, Gupta:2023evt,
DeRenzis:2024dvx } as it provides a fast, efficient resource to compute
estimates for parameter errors without the need for full Bayesian parameter
estimation. For a comprehensive overview of the caveats and limitations of the
Fisher-matrix formalism see Ref.~\cite{Vallisneri:2007ev}.

A fiducial, easily rescalable cosmic merger rate $\mathcal{R}$ of \num{$10^5\,
{\rm yr^{-1}}$}, chosen in agreement with the current BBH and BNS estimates
from LIGO/Virgo/KAGRA~\cite{LIGOScientific:2025pvj}, is used to compute the
expected observation time \tobs for a tuple (\nth, $X_{\rm th}$) under the
assumption that the mergers are Poisson-distributed in time; $X_{\rm th}$
represents a threshold in one of the key metrics. The computation of the
observation time is further bootstrapped \num{100}-fold
\cite{2020sdmm.book.....I} over a \num{3-year} period to mitigate
finite-sampling effects from the set of simulated binary merger; median values
are reported throughout this study. Statements on observation times beyond
\num{$3\,{\rm yr}$} are extrapolated and only reported up to \num{$5\,{\rm
yr}$}. The PBH binary merger rate was set to \num{$2\times10^6\,{\rm yr^{-1}}$}
consistent with the population model described below.

Sensitivity of the XG network to stochastic signals, such as the
gravitational-wave background (GWB) from the compact binary populations
presented here, is assessed by computing power-law integrated sensitivity
curves (PI curves)~\cite{Thrane:2013oya}.  A PI curve is the locus where an
envelope of stochastic signals each described via a power-law model with a
distinct index is tangent to the detector sensitivity, assuming a certain
detection statistic threshold (i.e., signal-to-noise ratio $\rho=3$). In
practice, this translates to the collection of points at which the tangent
power-law stochastic signal would satisfy a given $\rho$. The $\rho$ statistic
used here is defined for two detectors, $I$ and $J$, assuming joint observing
time $T_{\rm obs}$, as~\cite{Thrane:2013oya}
\begin{equation}
\rho_{IJ} = \sqrt{2\,T_{\rm obs}}\, \left[ \int {\rm d}f \,
\frac{\gamma^2_{IJ}(f) S^2_h(f)}{S^I_n(f) S^J_n((f)} \right]^{1/2} \,,
\label{eq:PI_tho}
\end{equation}
where $\gamma_{IJ}$ is the overlap reduction function~\cite{Romano:2016dpx} for
the two detectors, which quantifies the decoherence between the two detector
sites and orientations, while $S_h(f)$ and $S_n(f)$ represent the
gravitational-wave power spectral density of the stochastic background and
noise power spectral density of each detector, respectively.  The GWB SNR
$\rho_{IJ}$ scales with the square root of the observing time.

\subsection{Sources} \label{sec:simulations}

The \num{$10^6$} signals of all binary populations are generated using the
frequency-domain waveform models \texttt{IMRPhenomXHM}
\cite{Garcia-Quiros:2020qpx, Garcia-Quiros:2020qlt} for BBHs and PBH binaries
and \texttt{IMRPhenomXAS\_NRTidalv2}~\cite{Mihaylov:2021bpf} for BNS. The lower
frequency cutoff is set to \num{$f_l=2\,{\rm Hz}$} for the ET detectors,
\num{$f_l=3\,{\rm Hz}$} for LIGO-India, and \num{$f_l=5\,{\rm Hz}$} for CE,
while the upper frequency cutoff is set to \num{$f_u=\min\{1024\,{\rm Hz},
8\,f_{\rm isco}=8\,(6^{3/2}\pi\, GM/c^3)^{-1}\}$} for all detectors (where $G$
is Newton's constant and $c$ is the speed of light). The frequency step size is
set to \num{$\Delta f = 2^{-4}\,{\rm Hz}$}. The GW signal model parameters
consist of detector-frame chirp mass $M_c=\eta^{3/5}M$, symmetric mass ratio
$\eta$, spin parameter vectors $\vec{\chi}_1,\vec{\chi}_2$, luminosity distance
$D_L$, time $t_c$ and phase $\phi_c$ of coalescence, inclination angle $\iota$,
right ascension $\alpha$, declination $\delta$, and polarization angle $\psi$,
as well as the tidal deformability parameters $\Lambda_1$ and $\Lambda_2$ for
BNSs.

The BBH and BNS redshift distributions follow the merger rate density for field
BBHs from Ref.~\cite{Ng:2020qpk} with $z\in[0,10]$; for BNS, the
Madau-Dickinson~\cite{Madau:2014bja} star formation rate with $z\in[0,5]$ is
assumed, neglecting metallicity effects; more details on both poulations, e.g.
the applied time delay distributions, are reported in
Ref.~\cite{Borhanian:2022czq}. The BBH source-frame masses are distributed
according to the median Power-Law+Peak model from the Third Gravitational-Wave
Transient Catalog~\cite{Abbott:2020gyp}, while the BNS source-frame component
masses follow a Gaussian distribution $\mathcal{N}(\mu=1.35,\sigma=0.15)$
constrained within $[1,2]M_\odot$. The tidal deformability of the component
neutron stars is computed from their masses for the \texttt{SLY} equation of
state~\cite{Douchin:2001sv} as implemented in \texttt{bilby}
\cite{Ashton:2018jfp}.  The component spins are chosen to be aligned with the
orbital angular momentum of the binary system with uniformly distributed
projections, $\chi^{\rm BBH}_{i,z}\in[-0.75,0.75]$ and $\chi^{\rm
BNS}_{i,z}\in[-0.05,0.05]$, with $i=1,2$.
The luminosity distance $D_L$ is calculated from the sampled redshifts using
the \num{\texttt{Planck18}} cosmology~\cite{Planck:2018vyg} while both the time
and phase of coalescence are set to \num{$t_c=\phi_c=0$} for all signals. The
sampling method for the four angular parameters is the same for both
populations with $\cos(\iota),\, \cos(\pi/2-\delta)$ uniformly distributed in
$[-1,1]$ and $\alpha,\, \psi$ uniformly distributed in $[0,2\pi]$.
 
Further, a speculative population of stellar-mass PBHs is studied which may
have formed in the very early universe from, for example, the collapse of large
amplitude density perturbations (see e.g. Refs.~ \cite{Green:2024bam,
Escriva:2022duf, Suyama:2025dui} for recent reviews). If such a population
exists, since they would have populated the universe long before
matter-radiation equality and only interact gravitationally, they would
automatically explain some fraction of the dark matter energy density budget.
Their abundance, usually quantified by $f_{\rm PBH}=\Omega_{\rm
PBH}/\Omega_{\rm CDM}\leq1$, is the fraction of the dark matter energy density
today $\Omega_{\rm CDM}$ that is contained within PBHs. In the stellar-mass
range, GW observations place some of the most stringent constraints on their
abundance such that $f_{\rm
PBH}\lesssim10^{-3}$~\cite{Andres-Carcasona:2024wqk}. While initial PBH
clustering is expected to have a negligible impact on this mass
range~\cite{Crescimbeni:2025ywm}, extended mass distributions may relax the
constraints~\cite{Carr:2023tpt}.
However, even the detection of a single PBH binary would be extremely powerful
for narrowing down our theories of the early universe~\cite{Cole:2017gle}, as
well as explaining a fraction of the dark matter. Next-generation GW detectors
open up the possibility of detecting very high redshift BBH mergers, which
would provide smoking gun evidence of a primordial origin~\cite{Chen:2024dxh,
Iacovelli:2024aei, Martinovic:2020hru}.
 
Assuming a log-normal probability density function for the mass distribution of
PBHs, 
\begin{eqnarray}
\psi(m) = \frac{1}{m \sigma \sqrt{2 \pi}}\exp\left\{-\frac{[\log(m/ m_c) + 0.5
\sigma^2]^2} {2  \sigma^2}\right\}
\end{eqnarray}
centred on PBH mass $m_c=50\,M_\odot$ with width $\sigma=0.7$, the differential
merger rate is calculated as~\cite{Raidal:2018bbj}
\begin{eqnarray}
{\rm d} R(m_1,m_2,z)&=\frac{1.6\times10^6}{\rm Gpc^3\, yr} \,f_{\rm sup}
\,f_{\rm PBH}^{53/37}\left( \frac{m_1m_2} { M^2}\right)^{-34/37}
M^{-32/37}\nonumber\\ &\times \psi(m_1) \psi(m_2)(1 + z)^\alpha {\rm d}m_1 {\rm
d}m_2,
\end{eqnarray}
where $f_{\rm sup}$ is a suppression factor that quantifies the disruption of
binaries due to smaller PBHs before they merge. The fraction $f_{\rm
PBH}=1.2\times10^{-3}$ and the values of $m_c$ and $\sigma$ in the mass
distribution are taken so as to conservatively respect current observational
constraints. The redshift dependence is parametrized by a power-law where
$\alpha=1.28$ represents a PBH population which is Poisson distributed
\cite{Raidal:2017mfl, Raidal:2018bbj,Mukherjee:2021ags,Mukherjee:2021itf}.
To obtain binary PBH source population, the binary component masses $m_1,
m_2$ and the redshift $z$ are sampled from the differential merger rate between
$1-100\,{\rm M_\odot}$ and $1-100$, respectively, with fixed $f_{\rm sup}=1$.
All population-level results for $f_{\rm sup}=10^{-3}$ are obtained by
rescaling the results of this population.

Finally, the GWB signal is computed for all of the simulated source populations. The adimensional spectral
energy density $\Omega_{\rm GWB}(f) = \rho_c^{-1}\, {\rm d}\rho/{\rm d} \ln f$ is employed, in line with
similar works in the literature that aim to assess sensitivity to compact binary
backgrounds~\cite{Regimbau:2011rp, Callister:2020arv, Regimbau:2022mdu,
Branchesi:2023mws, Renzini:2024pxt, Ebersold:2024hgp}.  The spectrum for compact
binaries is defined as~\cite{Phinney:2001di}
\begin{equation} \Omega_{\rm GWB}^{\rm CBC} (f) = \frac{f}{\rho_c} \int {\rm d}z
\frac{R(z)}{(1+z)H(z)} \left\langle \, \frac{{\rm d}E_s}{{\rm d}f_s}\bigg|_{f_s} \right\rangle, 
\label{eqomegagw}\end{equation}
where $R(z)$ is the source-frame merger rate per comoving volume and ${\rm d}E_s/{\rm d}f_s$
is the binary source-frame energy spectrum, while $\rho_c$ is the critical
energy density of the Universe, and $H(z)$ is the Hubble parameter at redshift
$z$. The angle brackets indicate an ensemble average over the population
considered, which is performed here using the  {\tt
popstock}~\cite{Renzini:2024pxt} code.

\subsection{Detector networks} \label{sec:networks}

\begin{figure*}[t]
    \includegraphics[width=0.95\linewidth]{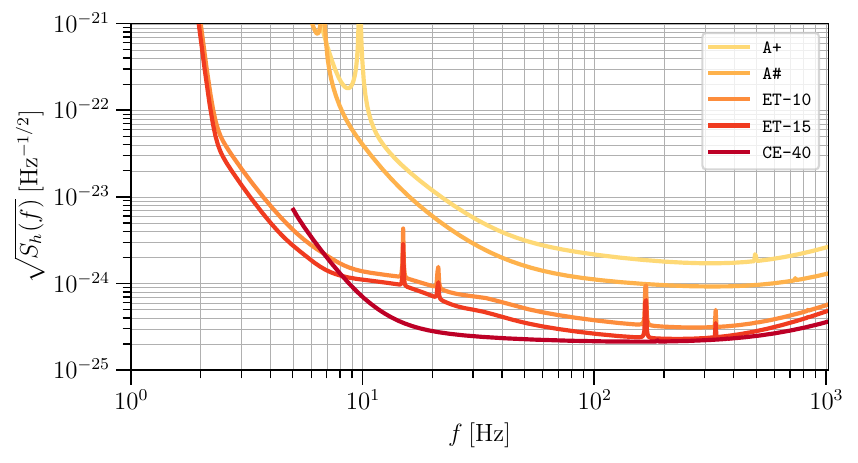}

    \caption{Detector sensitivity curves used in this study, see
    Sec.~\ref{sec:networks}. The \texttt{A+} and \texttt{A\#} curves are used
    for the LIGO-India detector \texttt{I}, \texttt{ET-10} and \texttt{ET-15} for
    ET configurations \ett and \etl, respectively, and \texttt{CE-40} for the CE
    facility \ce.}
    \label{fig:noise_curves}
\end{figure*}

This study considers \num{15} different XG detector networks, consisting of
\num{five} XG facility configurations, namely \ett, \etl, \ce, \ettc, and
\etlc, which are studied in isolation or in tandem with one of \num{two}
LIGO-India configurations, \texttt{I+} and \texttt{I\#}, using the A+ or A\#
detector sensitivities, respectively. The adopted detector sensitivity curves
are shown in Figure~\ref{fig:noise_curves} and crucially all detectors are
assumed to operate immediately at their respective design sensitivity. The
detector configurations are specified in \ref{app:detectors}. The following
labels are used throughout this study to refer to the networks, detectors, and
sensitivities:
\begin{itemize}
    \item \ett: single $10\,{\rm km}$, triangular interferometer at fiducial ET
    site in Sardinia, Italy~\cite{Abac:2025saz}, with \texttt{ET-10} ($10\,{\rm
    km}$-xylophone) sensitivity curve in \texttt{ET10kmcolumns.txt} from
    Ref.~\cite{et10km} for each of the three sub-interferometer;
    \item \etl: two $15\,{\rm km}$, L-shaped interferometers at fiducial ET
    sites in Sardinia, Italy and in the Meuse-Rhine region between the
    Netherlands, Belgium, and Germany~\cite{Abac:2025saz}, with \texttt{ET-15}
    ($15\,{\rm km}$-xylophone) sensitivity curve in \texttt{ET15kmcolumns.txt}
    from Ref.~\cite{et15km} for both interferometers, the two detectors are
    rotated by $45^\circ$ with respect to each other;
    \item \ce: single $40\,{\rm km}$, L-shaped interferometer at fiducial CE
    site in Idaho, USA~\cite{Reitze:2019iox}, with \texttt{CE-40} sensitivity
    curve in \texttt{cosmic\_explorer\_40km.txt} from Ref.~\cite{ce40km};
    \item \indp and \inds: single $4\,{\rm km}$, L-shaped interferometer at the
    LIGO-India site in Maharashtra, India~\cite{Unnikrishnan:2013qwa}, with
    \texttt{A+} sensitivity curve in \texttt{aplus.txt} from
    Ref.~\cite{apl:etd} or \texttt{A\#} sensitivity curve in
    \texttt{asharp\_strain.txt} from Ref.~\cite{ash}, respectively;
\end{itemize}

\noindent While this investigation focuses on LIGO-India, any other
current-generation detector---namely LIGO-Hanford, LIGO-Livingston, Virgo, or
KAGRA---would have a similar impact \cite{Vitale:2018nif} if operated at the
studied sensitivities in tandem with XG detectors. Regardless, with LIGO-India
under active construction and targeting observations to commence in 2030, this
facility presents a natural choice for this study.

\section{Results}
\label{sec:results}

\subsection{Next-generation timing} \label{sec:timing}

\begin{figure*}[p!] \centering
    \includegraphics[width=\linewidth]{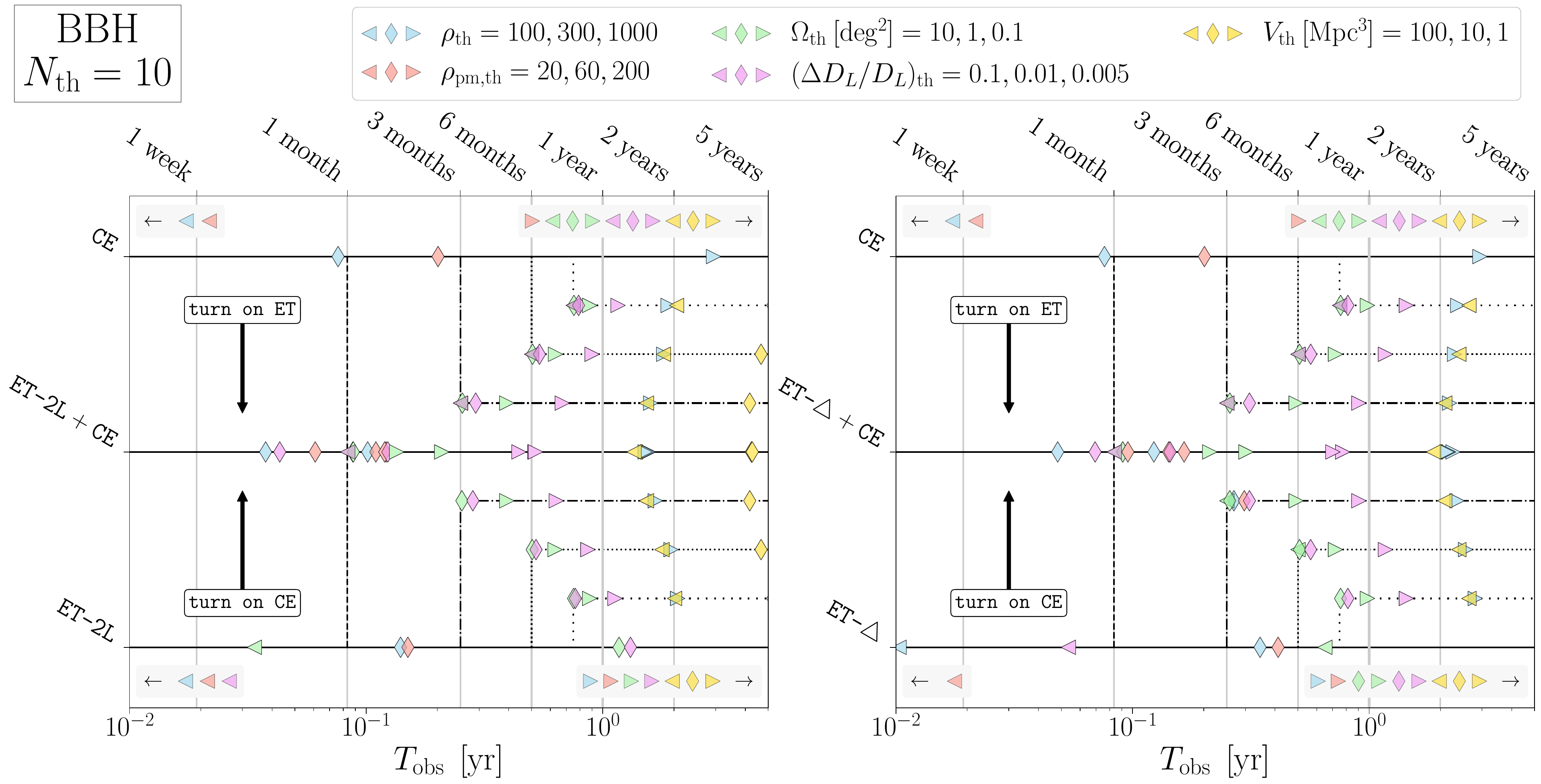}
    \includegraphics[width=\linewidth]{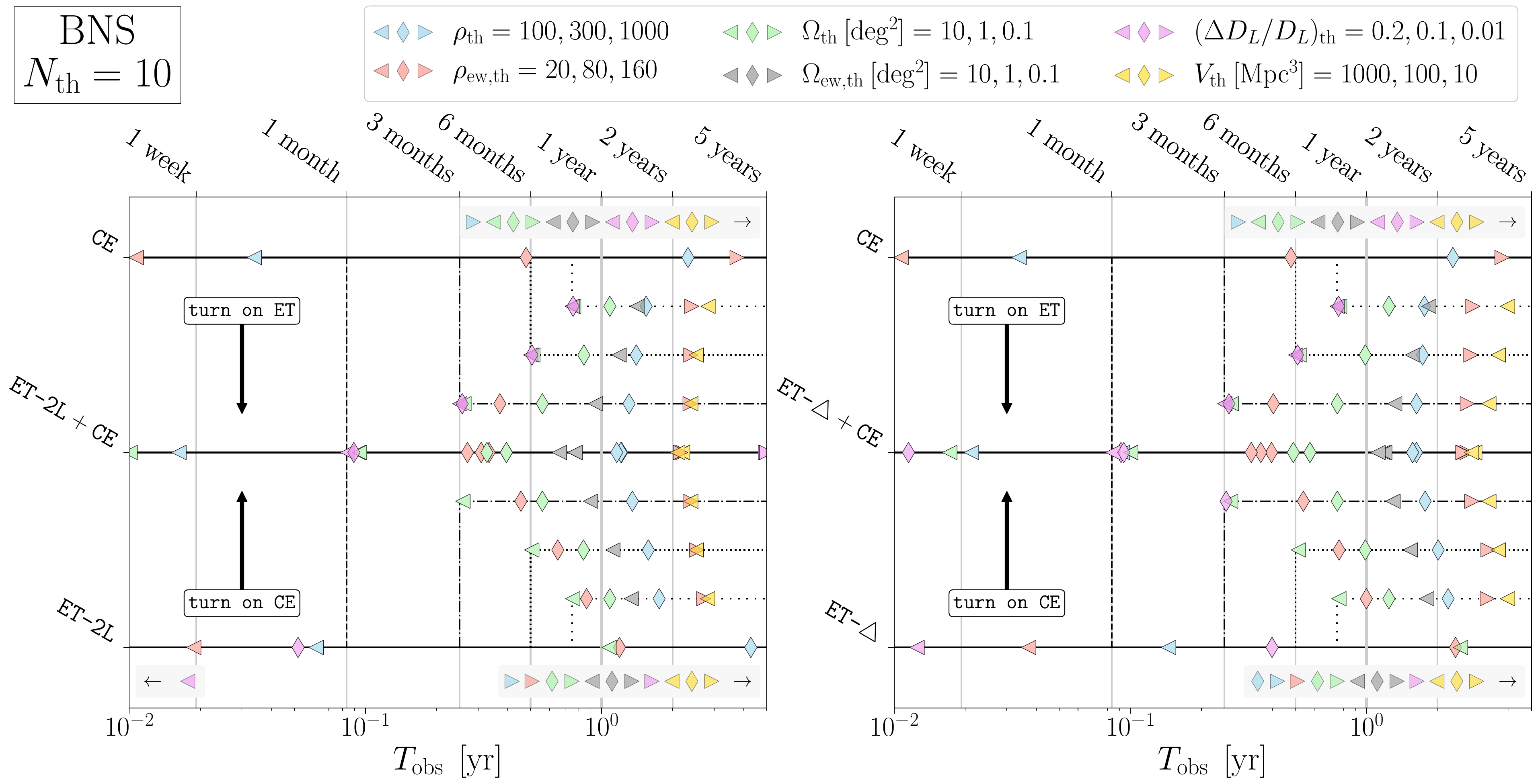}

    \caption{Comparison of the expected observation times \tobs to reach a
    target of \num{$\nth = 10$} BBH (top) and BNS (bottom) signals satisfying
    \num{three} different thresholds for SNR \snrth, sky area \skyth, relative
    luminosity distance error \disth, comoving error volume \volth, post-merger
    SNR \pmsnrth, early-warning SNR \ewsnrth, and sky area \ewskyth. For each
    of these, three thresholds are considered and indicated with different
    marker shapes.  The solid, horizontal lines represent the timelines for the
    base detectors (CE atop and ET beneath) and the combined network (center)
    to reach the targets; in addition to the impact of potential observation
    delays of {1 (dashed), 3 (dashed-dotted), 6 (dotted), and 9 (loosely
    dotted) months}. Targets that are either completed within \num{$0.01\,{\rm
    yr}$} or that cannot be reached within \num{$5\,{\rm yr}$} are places in
    grey boxes.
    For example, a single L-shaped \ce detector would not be able to detect
    \num{$\nth=10$} BBHs to within \num{$\skyth=10\,{\rm deg^2}$}
    (\textcolor{excol!60!white}{$\blacktriangleleft$}, top left panel) within
    $5\,{\rm yr}$, while \etl and the joint network \etlc would reach this
    target within \num{two weeks} or less than \num{$0.01\,{\rm yr}$},
    respectively.}

    \label{fig:timings_N_10}
\end{figure*}

To investigate the impact of timing and coordination of XG detector facilities,
particularly the potential for delays in observation campaigns,
Fig.~\ref{fig:timings_N_10} compares the expected observation times \tobs to
detect \num{$\nth=10$} BBH and BNS signals for metrics of interest (i) - (vii)
at \num{three} target thresholds for the \num{five} studied XG facility
configurations of ET and CE, \ett, \etl, \ce, \ettc, and \etlc.  In
particular, it illustrates the impact of either of the base XG facilities
experiencing long-term delays of \num{$\tau_{\rm delay}=1,3,6,9$ months} within
a \num{three}-year-long observation campaign.
\ref{app:tables} presents the respective values of the plotted observation
times. Metrics will be neglected in the discussion below if all networks
require observation times $\tobs\gtrsim5\,{\rm yr}$ to reach the targeted
thresholds, as potential errors in the extrapolation of the observation time
beyond \num{5} years would make these results unreliable. \ref{app:maps} presents
more complete maps of expected observation times \tobs over $\nth\in[1,100]$
for a wide range of targeted thresholds in the key metrics of interest in this
study.

Figure~\ref{fig:timings_N_10} presents the expected observation times \tobs as
markers along horizontal lines which represent different detector and network
configurations: solid lines signify the base XG configurations, \ett, \etl, \ce,
and the two networks, \ettc and \etlc, combining the ET configurations with CE.
Non-solid lines indicate the observation scenarios in which one of the  base XG
facilities experiences long-term delays of \num{$\tau_{\rm delay}=1,3,6,9\,{\rm
months}$} ahead of their observation campaign. The marker colors and shapes
indicate the metrics of interest and three different target thresholds,
respectively. In particular, each non-solid line begins at a specific time
from either the top (CE) or bottom (ET) solid line in each panel, indicating the
time CE or ET operates by themselves, before the other facility turns on after
the delay. Conversely markers will not be present along the non-solid
lines if the respective base XG facility reaches the targeted threshold before
the other facility turns on.

The SNR metrics---full signal, BBH post-merger, and BNS early-warning---are the
least sensitive in observation time to such delays and adhere to the most
sensitive, operating GW detector facility since SNRs add as sum of squares. All
base XG configurations observe the targeted number of \num{$\nth=10$} signals
satisfying \num{$\snrth\mbbh=100$, $\pmsnrth\mbbh=20$, $\snrth\mbns=50$, and
$\ewsnrth\mbns=20$} within \num{one} month, with \ett requiring \num{two} for
\num{$\snrth\mbns=50$}.  Thus, months-long observation delays will not affect
the realization of these targets; at least in a significant way. This statement
holds for BBHs observed with \ce and \etl as well, albeit in a weaker sense, for
the higher SNR thresholds of $\snrth\mbbh=300$ and $\pmsnrth\mbbh=60$: both
networks require \num{less than three months} of observation time to reach the
target; \ett would require up to \num{five months}. The largest impact is seen
for the small populations of high-SNR events ($\snrth\mbbh=1000$,
$\pmsnrth\mbbh=200$, $\snrth\mbns=200$, and $\ewsnrth\mbns=160$) which require
\num{$\tobs\gtrsim3\,{\rm yr}$} in all shown XG networks. The addition of the
second XG facility after \num{$\tau_{\rm delay}=0.75\,{\rm yr}$} results in the
smallest relative speed-ups for the observation time which vary across the these
high-SNR events in the following ranges: \num{$5.3-24$x} for \ett,
\num{$2.9-5.5$x} for \etl, and \num{$1.2-1.6$x} for \ce.

The localization metrics---sky area, relative luminosity distance error, and
comoving error volume---are very sensitive to the number of detectors,
particularly at non-colocated sites, in the network. Hence, both \ett\
(effectively two colocated detectors) and \etl (two non-colocated detectors)
outperform the more sensitive, one-detector facility of \ce for these metrics:
\ce alone cannot achieve any of the target localizations in \skyth and \disth\
within \num{5} years, see also \ref{app:maps}, while \etl (\ett) will see BBH
and BNS signals with $\skyth\mbbh=10\,{\rm deg^2}$, $\disth\mbbh=0.1$,
$\skyth\mbns=10\,{\rm deg^2}$, and $\disth\mbns=0.1$ approximately \num{ten}
times every \num{2 weeks (8 months), 16 hours (3 weeks), 1.1 years (2.5 years),
and 3 weeks (5 months)}, respectively. This is in line with previous studies
that underlined how a single XG interferometer alone would not meet many
science requirements~\cite{Branchesi:2023mws,Abac:2025saz}. The two-facility
networks, \ettc and \etlc, speed up the observation times by \num{one to three}
orders of magnitude, decreasing the observation times for the aforementioned
metrics to approximately \num{few hours, few hours, less than a week, and few
days}, respectively. In fact, in comparison to the one-facility configurations,
these networks will observe the target of 10 BBH and BNS signals satisfying
{any of the investigated sky and distance localization metrics} so rapidly that
the resulting observation times are fully governed by the raw, short
observation time of the two-facility network in addition to whatever delays are
present in their component facilities; in other words, observation delays are
essentially equivalent to network-wide interruptions for achieving the studied
targets for the localization metrics.  This statement softens as the targeted
number of signals \nth becomes of order unity, as statistical fluctuations
become more relevant.

The early-warning sky area \ewskyth for BNSs and comoving volume error \volth,
are particularly poorly constrained. The single-facility configurations, even
\etl\ with its two non-colocated detectors, are unable to observe $\nth=10$ BBH
and BNS signals satisfying any of the targeted thresholds for \ewskyth and
\volth within \num{5} years while the two-facility networks achieve the targets,
\num{$\ewskyth\mbns = 10\,{\rm deg^2}$, $\volth\mbbh = 100\:(10)\, {\rm Mpc^3}$,
and $\volth\mbns = 1000\,{\rm Mpc^3}$}, within \num{0.66 -- 1.1 years, 1.4 --
1.9 (4.3 with \etlc) years and 2.1 -- 2.8 years}, respectively. Again, delays in the
observation times are effectively equivalent to extensions of the observation
campaign. 

\begin{figure*}[t]
    \centering
    \includegraphics[width=0.49\linewidth]{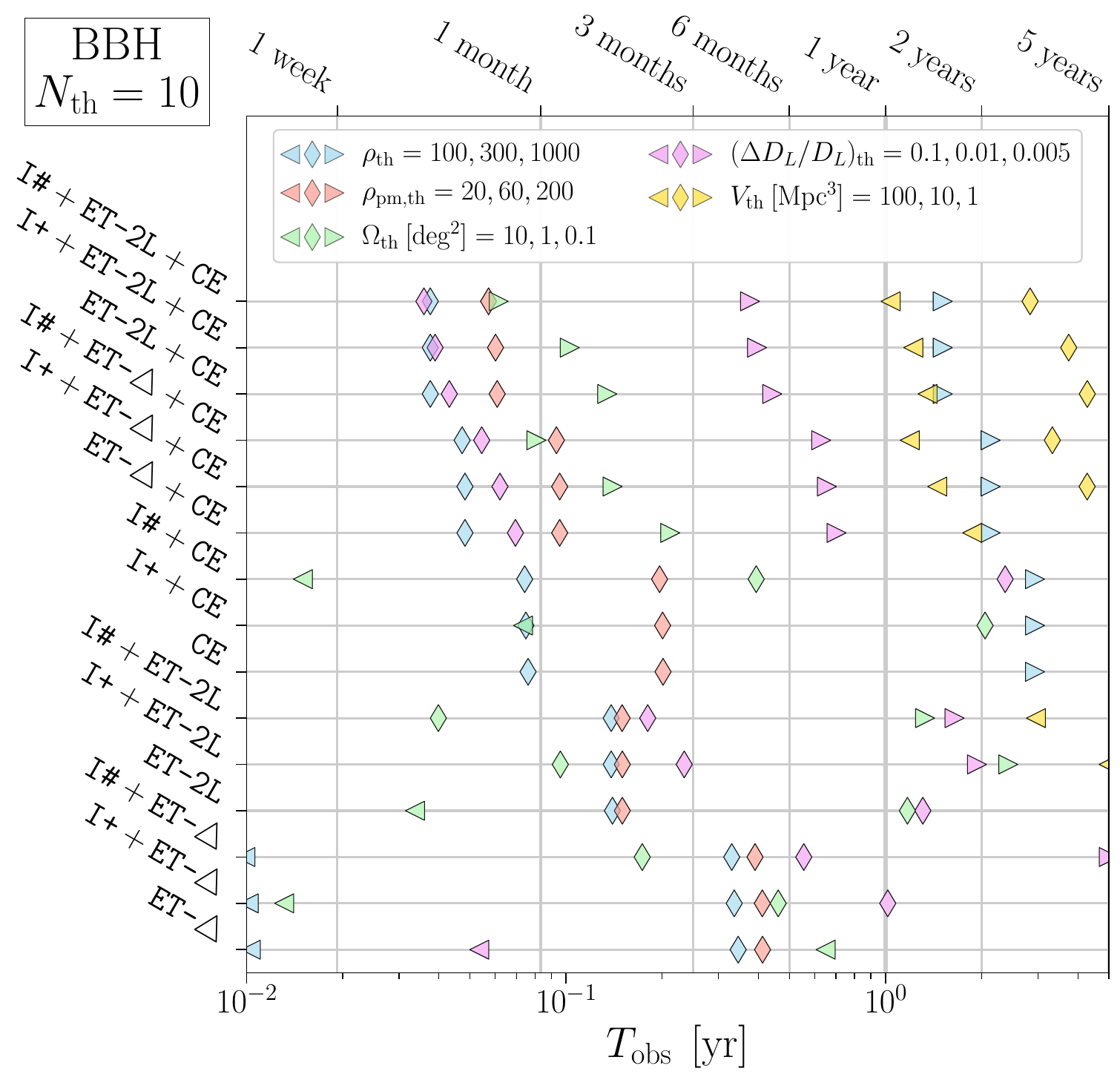}
    \includegraphics[width=0.49\linewidth]{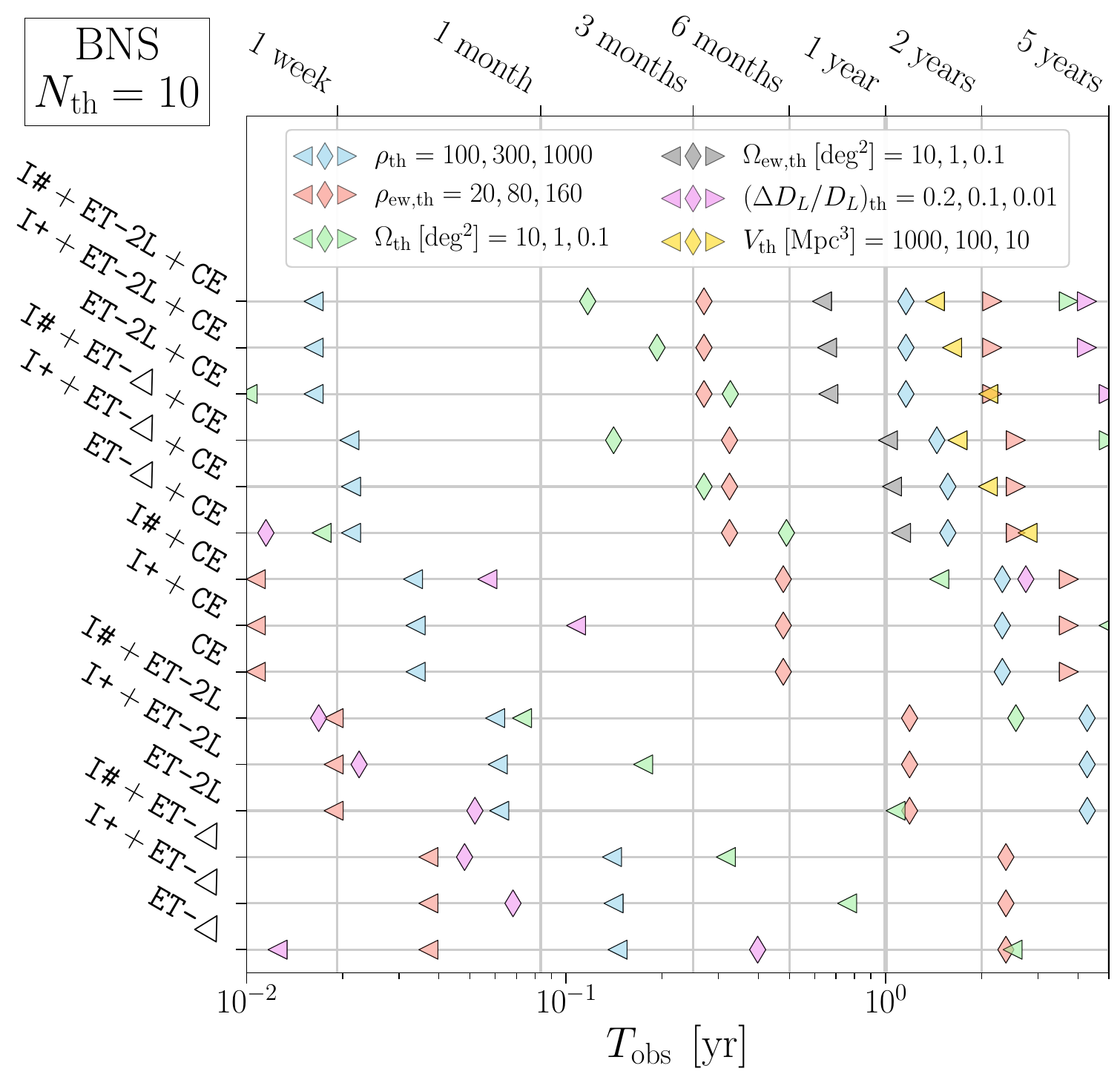}
    
    \caption{Impact of the addition of LIGO-India, either in \indp or \inds
    configuration, to \num{five} XG detectors and networks, \ett, \etl, \ce,
    \ettc, \etlc, on the expected observation times \tobs to reach a target of
    \num{$N_{\rm th} = 10$} BBH (left) and BNS (right) signals satisfying
    \num{three} different thresholds for SNR \snrth, sky area \skyth, relative
    luminosity distance error \disth, or comoving error volume \volth in
    addition to  post-merger SNR \pmsnrth for the BBHs and early-warning SNR
    \ewsnrth and sky area \ewskyth for the BNSs.}

    \label{fig:timings_N_10_india}
\end{figure*}

\subsection{Impact of LIGO-India} \label{sec:impact}

Figure~\ref{fig:timings_N_10_india} shows the impact of LIGO-India, in \indp or
\inds configuration, when operating in tandem with the \num{five} aforementioned
XG networks, \ett, \etl, \ce, \ettc, and \etlc. The respective data is collected
across Tables~\ref{tab:timings_bbh}, \ref{tab:timings_bns}, and
\ref{tab:india_bns}.  The addition of LIGO-India, even in the more sensitive
A\# configuration, has a negligible effect on the expected observation times
\tobs to reach the targeted thresholds for the SNR metrics which are dominated
by the more sensitive XG detectors in the network (see
Ref.~\cite{Vitale:2018nif} for a previous investigation). The same applies to
the early-warning sky area \ewskyth for BNS signals.

The other localization metrics paint the opposite picture: the addition of
LIGO-India dramatically decreases the observation times for the these metrics
in the single-facility configurations, \ett, \etl, and \ce. Both BBH and BNS
signals see \num{one to two} orders of magnitude speed-ups in the observation
times to achieve the targets for the sky area and luminosity distance error
metrics. In both cases, the A\# configuration roughly halves the observation
times compared to the \indp configuration. The impact is expectedly less severe
for the two-facility networks, \ettc and \etlc, and of the order of
\num{10-140\%} for BBH and \num{10-250\%} for BNS signals. Hence, the impact of
a current-generation detector operating in tandem with an XG network is
\emph{not} negligible in the context of reaching the investigated science
goals, particularly when potential delays in the observation time of XG
facilities are considered.

\begin{figure*}[t]
    \centering
    \includegraphics[width=\linewidth]{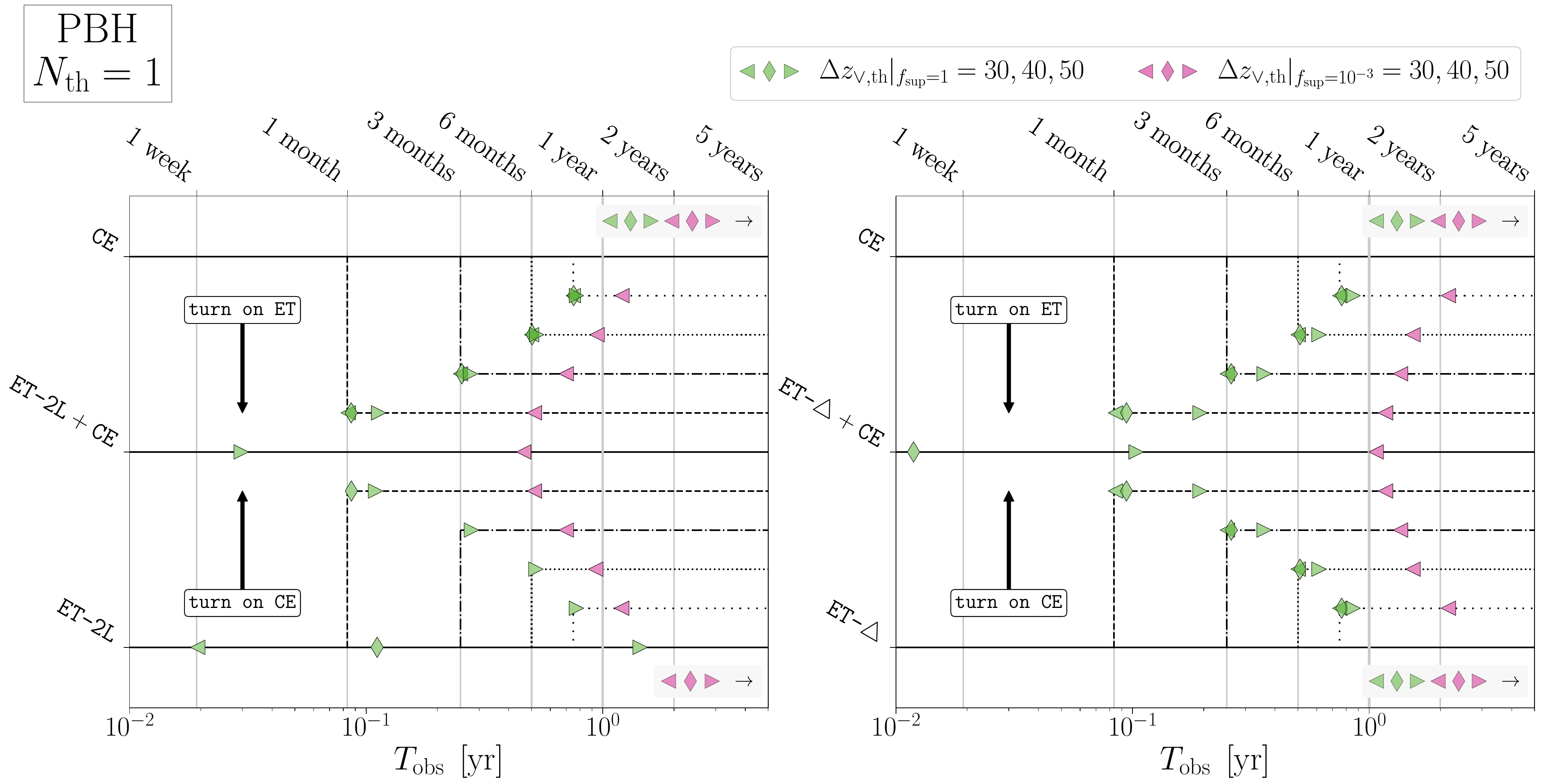}

    \caption{Comparison of the expected observation times \tobs to
    reach a target of \num{$\nth = 1$} PBH signal satisfying \num{three}
    different thresholds for the lower redshift error bound
    $\zloth\in\{30,40,50\}$ at two values for the PBH-formation suppression
    factor $f_{\rm sup}\in\{10^{-3},1\}$ when observed by the three base XG
    facility configurations, \ett (right), \etl (left), \ce (both), and two
    networks, \ettc, \etlc, combining the ET configurations with CE. The solid,
    horizontal lines represent the timelines for the base detectors (CE atop and
    ET beneath) and the combined network (center) to reach the targets. The
    non-solid lines take potential observation delays of {1 (dashed),
    3 (dashed-dotted), 6 (dotted), and 9 (loosely dotted) months} in
    either base detector into account, with the respective non-solid, vertical
    lines indicating the end of the delay. Grey boxes indicate---for each base
    detector---the targets that either are completed within \num{$0.01\,{\rm
    yr}$} or that cannot be reached within \num{$5\,{\rm yr}$.}}

    \label{fig:timings_N_1_pbh}
\end{figure*}

\subsection{Discernability of coalescences of primordial black holes} \label{sec:pbh}

The redshift at which stellar-mass BBHs merge is one of the key indicators to
discern primordial from astrophysical-origin BBHs~\cite{Mancarella:2023ehn}, as
the latter are not expected to merge at redshifts \num{$z\gtrsim30$}
\cite{Ng:2022agi,Stasenko:2024pzd}. Hence, Fig.~\ref{fig:timings_N_1_pbh}
presents the expected observation times \tobs for detecting at least
\num{$\nth=1$} PBH signal satisfying \num{$\zloth=30,40,50$} for the lower
redshift error bounds at two values for the suppression factor \num{$f_{\rm
sup}=\{10^{-3},1\}$}, which captures our uncertainty on how many PBHs that form
binaries in the early Universe are disrupted before merging.  In the
pessimistic scenario of $f_{\rm sup}=10^{-3}$, none of the base XG detector
facilities, \ett, \etl, and \ce, will be capable of observing even a single PBH
signal to be from $\zloth\ge30$ within $5\,{\rm yr}$, requiring joint
observation of this population with the ET-CE networks, \ettc and \etlc.
Excitingly, the results show that with both XG detectors online, a single PBH
binary with redshift $\zloth\geq30$ will be detectable \num{within 6 months} in
\etlc and \num{within a year} in \ettc. Therefore, any delay in the observation
campaign of either ET or CE would directly translate to an equal delay in
achieving a confident detection of a PBH binary based on redshift measurements
alone.  Even at $f_{\rm sup}=1$, only the two-detector facility \etl would be
capable of discerning, \num{almost once a week}, if a BBH is merging at
redshift $z\ge30$, and thus of primordial origin (see
also~\cite{Abac:2025saz} for similar conclusions). Figure~\ref{fig:pbh} in
\ref{app:maps} shows that LIGO-India in A\# configuration would enable the
triangular ET configuration \ett to observe \num{$\nth=1$} PBH merger with
\num{$\zloth=30$}, but the impact is much less pronounced than for the other
science metrics as the LIGO-India detector is not sensitive enough to detect
most of the very-high-redshift BBH signals. These results are expected
since the lower redshift error bound $\zloth=z(D_L - \Delta D_L)$ directly
depends on the luminosity distance error $\Delta D_L$ which, as a localization
metric, is very sensitive to the number of detectors in the network.

These results on the discernability of PBH binary mergers appear more
optimistic than what has been found in Ref.~\cite{Mancarella:2023ehn}. This seeming
discrepancy is a result of the assumptions made in the underlying PBH
mass distribution, favoring heavy binary systems with louder signals and
improved distance/redshift estimation.
 
\subsection{Sensitivity to power-law stochastic backgrounds} \label{sec:stochastic} 

\begin{figure} \centering
    \includegraphics[width=0.98\textwidth]{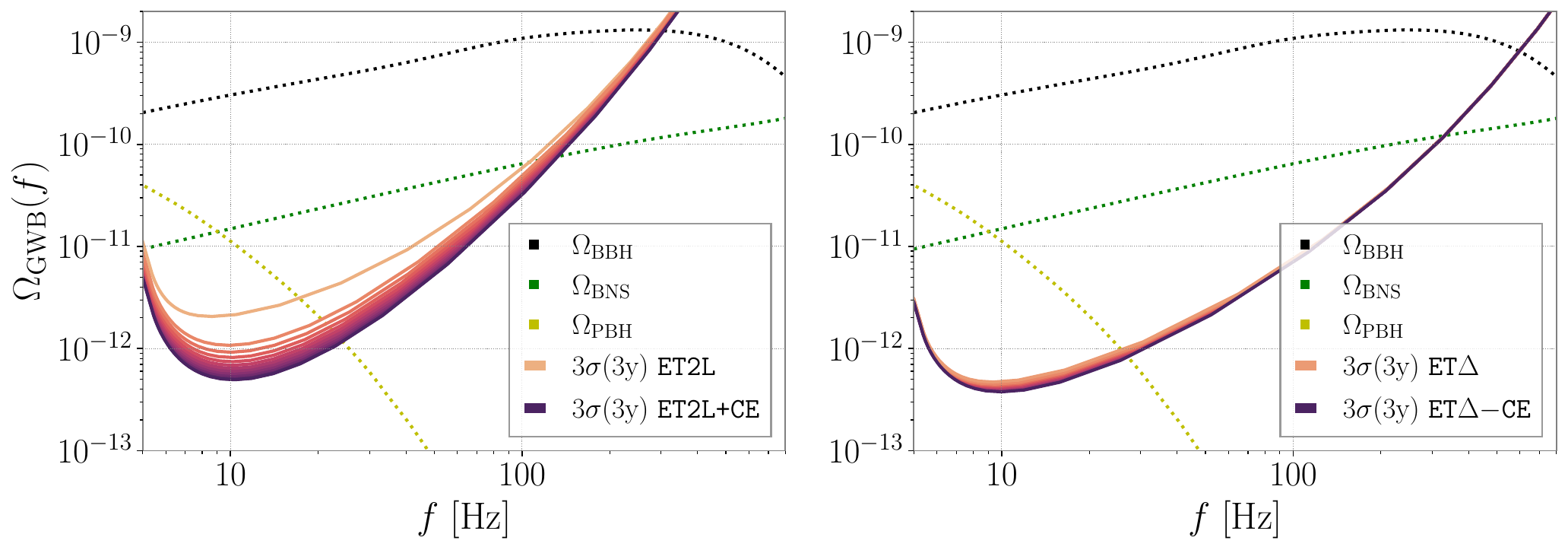}
    \\\medskip
    \includegraphics[width=0.98\textwidth]{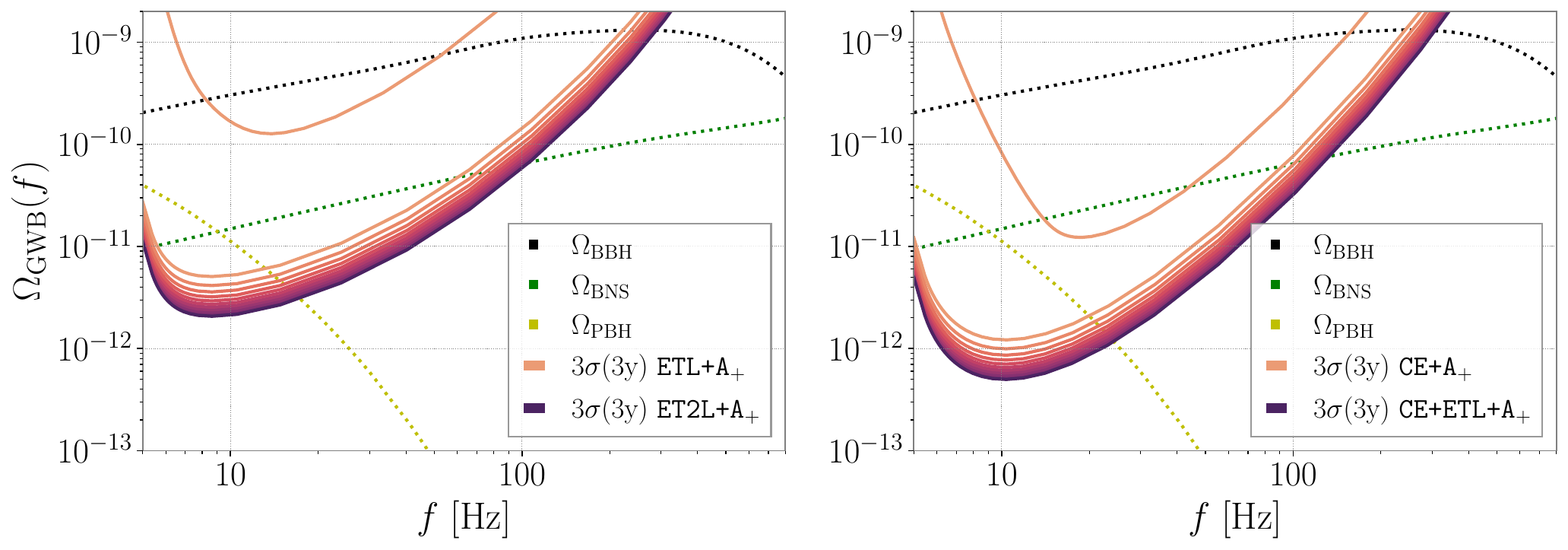}
    \caption{Power-law integrated (PI) curves for different XG networks, always
    considering a first set of detectors are online followed by a second
    detector with a delay of $3N$ months: the \etl detectors, followed by the
    \ce detector (top left); the \ett detector, followed by the \ce detector
    (top right); a single \etol detector and the \indp detector, followed by a
    second \etol (bottom left);  the \ce detector and the \indp detector,
    followed by the \etl detectors. In all cases, each curve represents a 3
    year observation where the first set of detectors are online for the first
    $3N$ months, while the full network is online for $3(12-N)$ months. The
    dotted lines represent the GWB spectral density $\Omega_{\rm GWB}^{\rm
    CBC}(f)$ calculated for each CBC population considered here.}
    \label{fig:stochastic_PI}
    \end{figure}

The following observing scenarios relevant to stochastic searches are
investigated:
\begin{itemize}
    \item the \etl detectors are online before the \ce detector (top left panel
    of Fig.~\ref{fig:stochastic_PI});
    \item the \ett detector is online before the \ce detector (top right panel
    of Fig.~\ref{fig:stochastic_PI});
    \item a single \etol detector and the \indp detector are online, followed
    by a second \etol (bottom left panel of Fig.~\ref{fig:stochastic_PI});
    \item the \ce detector and the \indp detector are online, followed by the
    \etl detectors (bottom right panel of Fig.~\ref{fig:stochastic_PI}).
\end{itemize}
In all these cases, it is assumed that the second detector comes online with a
delay of $3N$ months and the PI curve is re-computed for each $N$, with $N$
spanning $0-12$. This implies each curve should be interpreted as a total
observing time of 3 years, where only a sub-set of detectors were online for
the first $3N$ months, while the full network is online for $3(12-N)$ months. 

The first two cases considered highlight the impact of observing a stochastic
signal with two or more XG detectors. Since stochastic searches require at
least two detectors with (approximately) uncorrelated noise to construct the
cross-correlation statistic \cite{Allen:1997ad,Romano:2016dpx,Renzini:2023qtj},
one cannot study the impact of having a single detector as this would require a
significantly different statistic, incompatible with the definition of PI
curves used here [Eq.~(\ref{eq:PI_tho})]; therefore, the case of a single CE
detector observing before being joined by the ET detectors cannot be studied.
Here, \etl is considered to consist of two detectors with uncorrelated noise,
while \ett to be comprised of two co-located independent detectors with
uncorrelated noise. The latter is an optimistic approximation inspired by
Ref.~\cite{Muratore:2022nbh} and commonly used in the
literature~\cite{Branchesi:2023mws}. 

In the case of \etl, the addition of CE to the network moderately impacts the
sensitivity to a stochastic signal; if CE comes online within an observing time
of 3 years, the final sensitivity can be impacted up to a factor of $\sim$5, as
shown in the top left panel of Fig.~\ref{fig:stochastic_PI}.  Conversely, in
the case of \ett considered here, CE would have a weak impact on the
observation, as seen in the top right panel of Fig.~\ref{fig:stochastic_PI}.
This difference is due to the fact that the suppression of the signal due to
the overlap reduction function $\gamma_{IJ}$ [cf. Eq.~(\ref{eq:PI_tho})] is
proportional to the distance between detectors, implying that co-located yet
uncorrelated detectors provide the ideal observing strategy for backgrounds. 

The second set of cases considered compare the performance of a hybrid network
where an XG detector is initially observing coherently with a
current-generation detector, and subsequently a second XG detector joins the
observing campaign. As in the studies presented above, LIGO India is considered
for this purpose. After verifying  that, in the configurations considered,
there is no significant difference between assuming \indp versus \inds
sensitivity, only the results for \indp are presented.

As shown in the bottom panels of Fig.~\ref{fig:stochastic_PI}, adding a second
XG detector to the network has a very significant impact to the network
stochastic sensitivity. In the first case, adding a second \etol to the \etol
$-$ \indp network increases the sensitivity by a factor of $\sim50$ almost
instantly, as seen in the left panel. In the second case considered, adding
\etl to the \ce+\indp network increases the sensitivity by an order of
magnitude. Similarly to directional studies discussed above, a network with at
least two XG detectors is a substantially different network to one with a
single XG, even when operating jointly with current-generation detectors. 

For reference, Fig.~\ref{fig:stochastic_PI} includes the total GWB signals from
the three populations considered in this work, $\Omega_{\rm GWB}^{\rm BBH}(f)$,
$\Omega_{\rm GWB}^{\rm BNS}(f)$, $\Omega_{\rm GWB}^{\rm PBH}(f)$, calculated as
in Eq.~(\ref{eqomegagw}). Note that a large portion of the signals that make up
these background spectra will be individually resolvable, hence in a realistic
detection scenario, the relevant stochastic portion of these signals may be the
result of signal subtraction or simultaneously estimated with resolved
components, as discussed
in~\cite{Regimbau:2016ike,Sachdev:2020bkk,Martinovic:2020hru,Biscoveanu:2020gds,
Zhou:2022otw,Zhou:2022nmt,Zhong:2024dss,Belgacem:2024ntv,Zhong:2025qno}.
Nevertheless, this study points out that all three of these spectra appear
significantly detectable, in the case where at least two XG detectors are
cross-correlated, and population characterisation studies employing the
stochastic signal only may be performed, as suggested in
Ref.~\cite{Giarda:2025ouf}.  On the other hand, as seen in
Fig.~\ref{fig:stochastic_PI}, a single XG detector correlated with a 2G
detector is likely to detect the louder BBH background but would not be able to
measure the neutron star and primordial signals.

\section{Conclusions} \label{sec:conclusion}

In summary, purely SNR-dependent quantities are not affected, or only very
marginally, by the timings of next-generation GW detector facilities within a
network. A single XG facility will quickly deliver most of the science output
that revolves around those targets. This includes studies aimed at constraining
the observed populations of compact binaries \cite{DeRenzis:2024dvx}, as well
as testing GR via e.g. black-hole spectroscopy \cite{Bhagwat:2023jwv,
Abac:2025saz, Berti:2025hly}.
On the other hand, long-term delays in or asynchronous operation of the
component facilities should be expected to significantly impact observation
campaigns for any quantities in GW astronomy that strongly correlate or depend
on the localization metrics, sky area and luminosity distance error. Crucially,
this includes all multi-messenger applications, although modern, innovative
optical telescopes such as the Vera C. Rubin Observatory could help mitigate
the impact of delays to some degree, see Ref.~\cite{Loffredo:2024gmx}.
In these cases, the observation times \tobs for networks containing both ET and
CE configurations are orders of magnitude shorter than for networks with only
one XG facility and any delay in the observation campaign of one of the
component facilities is equivalent to a network-wide interruption for these
metrics. Moreover, some configurations, notably those consisting of a single XG
detector alone, would not be able to meet some of the related science
requirements.

The addition of LIGO-India, in either the \indp or \inds configuration, while
distinctly less sensitive than the XG facilities, presents a significant
opportunity for softening the impact of such delays in the XG detector
facilities, particularly for the strongly affected localization metrics. As
expected and well understood~\cite{Vitale:2018nif}, the addition of LIGO-India
in either configuration would have negligible impact on the SNR metrics which
will be dominated by the XG facilities. These conclusions would likely hold for
any other of the existing detectors, if kept online while the XG networks
commence observations and if operated at the investigated sensitivities.

Only primordial stellar-mass black holes are expected to merge at redshifts
\num{$z\gtrsim30$} and thus a confident detection at redshifts beyond that
threshold would serve as evidence for primordial black holes. The results of
this study show that this metric requires at least two XG detectors in the
network, \etl\, \ettc, or \etlc, to confidently observe and discern that BBH
merger events originate from redshifts $z\ge30$. The addition of LIGO-India in
A\# configuration would only enable the \ett configuration to observe a few PBH
merger events at such high redshifts. However, even the detection of a single
PBH binary would bring about a paradigm shift in the fields of dark matter and
early-universe cosmology.

Finally, the impact of different hybrid networks on the detectability of a
stochastic signal was assessed. In particular it was shown that having at least
two XG facilities online at the same time considerably accelerates detection
and characterisation of GW backgrounds, including signals from different
compact binary populations.

This study has several limitations. First and foremost, the Fisher information
formalism is not as robust as full Bayesian parameter estimation and therefore
single-event statistics can be misleading. This has been mitigated by focusing
on a target of $\nth=10$ signals fulfilling the thresholds per metric and
further through bootstrapping the simulated populations. Further, this study is
limited to SNR and localization dependent metrics, discarding intrinsic GW
parameters such as black hole or neutron star masses and spins, spin
precession, orbital eccentricity, or neutron star tidal deformability, but also
extrinsic phenomenologies concerning cosmology and astrophysical populations
\cite{DeRenzis:2024dvx}. Additionally, the investigation of facility-wide
timing delays does not take into account that the multi-detector ET facilities
\ett and \etl could experience such a delay in a single component detector
only; it is implicitly assumed that the ET commissioning bodies will ensure
simultaneous operation of all component detectors. Finally, it is assumed that
all facilities under consideration commence observations at their design or
target sensitivities. While this is common practice for forecasting studies, it
is more relevant in the presented context of timing delays between facilities
since the application of very optimistic sensitivity curves from the onset of
observations could exacerbate the impact of such delays, as can be seen in
Ref.~\cite{Maggiore:2024cwf}. 

Taken together, the results of this paper show that a single XG facility can
indeed deliver a portion of the targeted science. However, some of the crucial
elements of the XG science case require both ET and CE. Cooperation is the real
key to success for GW astronomy in the coming decades.

\ack
We thank Rodrigo Tenorio and Riccardo Buscicchio for  discussions.  S.B., P.C.,
A.R., C.P., and D.G. are supported by ERC Starting Grant No.~945155--GWmining,
Cariplo Foundation Grant No.~2021-0555, MUR PRIN Grant No.~2022-Z9X4XS,
Italian-French University (UIF/UFI) Grant No.~2025-C3-386, MUR Grant ``Progetto
Dipartimenti di Eccellenza 2023-2027'' (BiCoQ), and the ICSC National Research
Centre funded by NextGenerationEU.
A.R. and D.G. are supported by MSCA Fellowships No.~101064542--StochRewind.
D.G. is supported by MSCA Fellowship No.~101149270--ProtoBH and MUR Young
Researchers Grant No. SOE2024-0000125.
M.M. is supported by the French government under the France 2030 investment
plan, as part of the Initiative d'Excellence d'Aix-Marseille Universit\'e --
A*MIDEX AMX-22-CEI-02.
Computational work was performed at CINECA with allocations through INFN and
the University of Milano-Bicocca, and at NVIDIA with allocations through the
Academic Grant program.

\section*{ORCID IDs}

Ssohrab Borhanian \orcidlink{0000-0003-0161-6109}
\href{https://orcid.org/0000-0003-0161-6109}{0000-0003-0161-6109},\\ Philippa
Cole \orcidlink{0000-0001-6045-6358}
\href{https://orcid.org/0000-0001-6045-6358}{0000-0001-6045-6358},\\ Arianna
Renzini \orcidlink{0000-0002-4589-3987}
\href{https://orcid.org/0000-0002-4589-3987}{0000-0002-4589-3987},\\ Costantino
Pacilio \orcidlink{0000-0002-8140-4992}
\href{https://orcid.org/0000-0002-8140-4992}{0000-0002-8140-4992},\\ Michele
Mancarella \orcidlink{0000-0002-0675-508X}
\href{https://orcid.org/0000-0002-0675-508X}{0000-0002-0675-508X},\\ Davide
Gerosa \orcidlink{0000-0002-0933-3579}
\href{https://orcid.org/0000-0002-0933-3579}{0000-0002-0933-3579}.

\appendix

\section{Detector specifications} \label{app:detectors}

The exact detector specifications can found in the dictionary at
\url{https://gitlab.com/sborhanian/gwbench/-/blob/master/gwbench/utils.py?ref_type=heads#L81}
with the following keys used for the various studied detector facilities:
\begin{itemize}
    \item \textsc{`ETS1'}, \textsc{`ETS2'}, \textsc{`ETS3'} for the \ett,
    \item \textsc{`ETN'} and \textsc{`ETS'} for \etl,
    \item \textsc{`CEA'} for \ce,
    \item \textsc{`I'} for \indp and \inds.
\end{itemize}

\section{Accompanying tables to the figures} \label{app:tables}

Tables~\ref{tab:timings_bbh}, \ref{tab:timings_bns}, and \ref{tab:india_bns}
collect the expected observation times \tobs represented by markers across
Figs.~\ref{fig:timings_N_10} and \ref{fig:timings_N_10_india} for all studied
detector and network configurations, as well as the targeted thresholds for the
SNR, sky area, luminosity distance error, comoving error volume, post-merger
SNR, early-warning SNR, and early-warning sky area metrics.

\begin{table*}[p]

\caption{Companion table for estimated observation times \tobs for BBHs collecting the same information as shown across the top row of Fig.~\ref{fig:timings_N_10} and the left panel of Fig.~\ref{fig:timings_N_10_india}. Each entry in the table corresponds to a marker in the figures for detector and network configurations \ett, \ettp, \etts, \ettc, \etl, \etlp, \etls, \etlc, \ce, \cep, and \ces.}

\begin{center}
\resizebox{0.63\textwidth}{!}{
\begin{tabular}{l|ccc|ccc|ccc}
\hline
\hline
\multirow{2}{*}{$\bm{T_{\rm obs}\mbbh \:\:[{\rm yr}]}$} & \multicolumn{3}{c|}{\snrth} & \multicolumn{3}{c|}{$\skyth \:\: [{\rm deg^2}]$} & \multicolumn{3}{c}{\disth} \\ 
 & 100 & 300 & 1000 & 10 & 1 & 0.1 & 0.1 & 0.01 & 0.005 \\ 
\hline
\hline
\etts  & 0.0099    & 0.33      & >5        & 0.0057    & 0.17      & >5        & 0.00087   & 0.55      & 5        \\ 
\ettp  & 0.01      & 0.34      & >5        & 0.013     & 0.46      & >5        & 0.0013    & 1         & >5       \\ 
\hline
\ett   & 0.01      & 0.35      & >5        & 0.65      & >5        & >5        & 0.054     & >5        & >5       \\ 
\hline
\ettc $\:(f_{\rm gap} = 0.75)$ &  &  & 2.8       &  & 0.76      & 0.98      &  & 0.81      & 1.4      \\ 
\ettc $\:(f_{\rm gap} = 0.5)$ &  &  & 2.6       & 0.5       & 0.51      & 0.72      &  & 0.57      & 1.2      \\ 
\ettc $\:(f_{\rm gap} = 0.25)$ &  & 0.26      & 2.4       & 0.25      & 0.26      & 0.49      &  & 0.31      & 0.9      \\ 
\ettc $\:(f_{\rm gap} = 0.083)$ &  & 0.12      & 2.3       & 0.084     & 0.091     & 0.3       &  & 0.14      & 0.77     \\ 
\ettc  & 0.002     & 0.048     & 2.1       & 0.00044   & 0.0077    & 0.21      & 0.0003    & 0.069     & 0.7      \\ 
\hline
\hline
\etls  & 0.005     & 0.14      & >5        & 0.0018    & 0.04      & 1.3       & 0.00044   & 0.18      & 1.6      \\ 
\etlp  & 0.0051    & 0.14      & >5        & 0.004     & 0.096     & 2.4       & 0.00054   & 0.23      & 1.9      \\ 
\hline
\etl   & 0.0051    & 0.14      & >5        & 0.034     & 1.2       & >5        & 0.0018    & 1.3       & >5       \\ 
\hline
\etlc $\:(f_{\rm gap} = 0.75)$ &  &  & 2.1       &  & 0.75      & 0.88      &  & 0.77      & 1.1      \\ 
\etlc $\:(f_{\rm gap} = 0.5)$ &  &  & 1.9       &  & 0.5       & 0.63      &  & 0.52      & 0.87     \\ 
\etlc $\:(f_{\rm gap} = 0.25)$ &  &  & 1.7       &  & 0.25      & 0.39      &  & 0.28      & 0.64     \\ 
\etlc $\:(f_{\rm gap} = 0.083)$ &  & 0.1       & 1.6       &  & 0.088     & 0.21      &  & 0.12      & 0.52     \\ 
\etlc  & 0.0016    & 0.038     & 1.5       & 0.00033   & 0.005     & 0.13      & 0.00023   & 0.043     & 0.44     \\ 
\hline
\hline
\ces   & 0.0029    & 0.074     & 2.9       & 0.015     & 0.39      & >5        & 0.0043    & 2.4       & >5       \\ 
\cep   & 0.0029    & 0.075     & 2.9       & 0.073     & 2         & >5        & 0.0088    & >5        & >5       \\ 
\hline
\ce    & 0.0029    & 0.076     & 2.9       & >5        & >5        & >5        & >5        & >5        & >5       \\ 
\hline
\cet $\:(f_{\rm gap} = 0.75)$ &  &  & 2.4       & 0.75      & 0.76      & 0.98      & 0.75      & 0.81      & 1.4      \\ 
\cet $\:(f_{\rm gap} = 0.5)$ &  &  & 2.3       & 0.5       & 0.51      & 0.72      & 0.5       & 0.57      & 1.2      \\ 
\cet $\:(f_{\rm gap} = 0.25)$ &  &  & 2.2       & 0.25      & 0.26      & 0.49      & 0.25      & 0.31      & 0.9      \\ 
\cet $\:(f_{\rm gap} = 0.083)$ &  &  & 2.2       & 0.084     & 0.091     & 0.3       & 0.084     & 0.14      & 0.77     \\ 
\cet   & 0.002     & 0.048     & 2.1       & 0.00044   & 0.0077    & 0.21      & 0.0003    & 0.069     & 0.7      \\ 
\hline
\cel $\:(f_{\rm gap} = 0.75)$ &  &  & 1.9       & 0.75      & 0.75      & 0.88      & 0.75      & 0.79      & 1.2      \\ 
\cel $\:(f_{\rm gap} = 0.5)$ &  &  & 1.8       & 0.5       & 0.5       & 0.63      & 0.5       & 0.54      & 0.9      \\ 
\cel $\:(f_{\rm gap} = 0.25)$ &  &  & 1.6       & 0.25      & 0.25      & 0.39      & 0.25      & 0.29      & 0.67     \\ 
\cel $\:(f_{\rm gap} = 0.083)$ &  &  & 1.5       & 0.084     & 0.088     & 0.21      & 0.084     & 0.12      & 0.52     \\ 
\cel   & 0.0016    & 0.038     & 1.5       & 0.00033   & 0.005     & 0.13      & 0.00023   & 0.043     & 0.44     \\ 
\hline
\hline
\hline
 & \multicolumn{3}{c|}{\pmsnrth} & \multicolumn{3}{c|}{} & \multicolumn{3}{c}{$\volth \:\: [{\rm Mpc^3}]$} \\ 
 & 20 & 60 & 200 &  &  &  & $100$ & $10$ & $1$ \\ 
\hline
\hline
\etts  & 0.0085    & 0.39      & >5        &  &  &  & >5        & >5        & >5       \\ 
\ettp  & 0.0087    & 0.41      & >5        &  &  &  & >5        & >5        & >5       \\ 
\hline
\ett   & 0.0088    & 0.41      & >5        &  &  &  & >5        & >5        & >5       \\ 
\hline
\ettc $\:(f_{\rm gap} = 0.75)$ &  &  & >5        &  &  &  & 2.6       & >5        & >5       \\ 
\ettc $\:(f_{\rm gap} = 0.5)$ &  &  & >5        &  &  &  & 2.4       & >5        & >5       \\ 
\ettc $\:(f_{\rm gap} = 0.25)$ &  & 0.3       & >5        &  &  &  & 2.1       & >5        & >5       \\ 
\ettc $\:(f_{\rm gap} = 0.083)$ &  & 0.17      & >5        &  &  &  & 1.9       & >5        & >5       \\ 
\ettc  & 0.0029    & 0.096     & >5        &  &  &  & 1.9       & >5        & >5       \\ 
\hline
\hline
\etls  & 0.0043    & 0.15      & >5        &  &  &  & 3         & >5        & >5       \\ 
\etlp  & 0.0045    & 0.15      & >5        &  &  &  & 5         & >5        & >5       \\ 
\hline
\etl   & 0.0045    & 0.15      & >5        &  &  &  & >5        & >5        & >5       \\ 
\hline
\etlc $\:(f_{\rm gap} = 0.75)$ &  &  & >5        &  &  &  & 2         & >5        & >5       \\ 
\etlc $\:(f_{\rm gap} = 0.5)$ &  &  & >5        &  &  &  & 1.8       & 4.7       & >5       \\ 
\etlc $\:(f_{\rm gap} = 0.25)$ &  &  & >5        &  &  &  & 1.5       & 4.2       & >5       \\ 
\etlc $\:(f_{\rm gap} = 0.083)$ &  & 0.11      & >5        &  &  &  & 1.4       & 4.2       & >5       \\ 
\etlc  & 0.0022    & 0.061     & >5        &  &  &  & 1.4       & 4.3       & >5       \\ 
\hline
\hline
\ces   & 0.0053    & 0.2       & >5        &  &  &  & >5        & >5        & >5       \\ 
\cep   & 0.0057    & 0.2       & >5        &  &  &  & >5        & >5        & >5       \\ 
\hline
\ce    & 0.0057    & 0.2       & >5        &  &  &  & >5        & >5        & >5       \\ 
\hline
\cet $\:(f_{\rm gap} = 0.75)$ &  &  & >5        &  &  &  & 2.6       & >5        & >5       \\ 
\cet $\:(f_{\rm gap} = 0.5)$ &  &  & >5        &  &  &  & 2.4       & >5        & >5       \\ 
\cet $\:(f_{\rm gap} = 0.25)$ &  &  & >5        &  &  &  & 2.1       & >5        & >5       \\ 
\cet $\:(f_{\rm gap} = 0.083)$ &  & 0.14      & >5        &  &  &  & 1.9       & >5        & >5       \\ 
\cet   & 0.0029    & 0.096     & >5        &  &  &  & 1.9       & >5        & >5       \\ 
\hline
\cel $\:(f_{\rm gap} = 0.75)$ &  &  & >5        &  &  &  & 2         & >5        & >5       \\ 
\cel $\:(f_{\rm gap} = 0.5)$ &  &  & >5        &  &  &  & 1.8       & 4.7       & >5       \\ 
\cel $\:(f_{\rm gap} = 0.25)$ &  &  & >5        &  &  &  & 1.5       & 4.2       & >5       \\ 
\cel $\:(f_{\rm gap} = 0.083)$ &  & 0.12      & >5        &  &  &  & 1.4       & 4.2       & >5       \\ 
\cel   & 0.0022    & 0.061     & >5        &  &  &  & 1.4       & 4.3       & >5       \\ 
\hline
\hline
\end{tabular}
\label{tab:timings_bbh}
}
\end{center}
\end{table*}

\begin{table*}[]

\caption{Companion table for estimated observation times \tobs for BNSs collecting the same information as shown across the bottom row of Fig.~\ref{fig:timings_N_10} and the right panel of Fig.~\ref{fig:timings_N_10_india}. Each entry in the table corresponds to a marker in the figures for detector and network configurations \ett, \ettp, \etts, \ettc, \etl, \etlp, \etls, \etlc, \ce, \cep, and \ces.}

\begin{center}
\resizebox{0.63\textwidth}{!}{
\begin{tabular}{l|ccc|ccc|ccc}
\hline
\hline
\multirow{2}{*}{$\bm{T_{\rm obs}\mbns \:\:[{\rm yr}]}$} & \multicolumn{3}{c|}{\snrth} & \multicolumn{3}{c|}{$\skyth \:\: [{\rm deg^2}]$} & \multicolumn{3}{c}{\disth} \\ 
 & 50 & 200 & 400 & 10 & 1 & 0.1 & 0.2 & 0.1 & 0.01 \\ 
\hline
\hline
\etts  & 0.14      & >5        & >5        & 0.32      & >5        & >5        & 0.0033    & 0.048     & >5       \\ 
\ettp  & 0.14      & >5        & >5        & 0.76      & >5        & >5        & 0.0038    & 0.068     & >5       \\ 
\hline
\ett   & 0.15      & >5        & >5        & 2.5       & >5        & >5        & 0.013     & 0.4       & >5       \\ 
\hline
\ettc $\:(f_{\rm gap} = 0.75)$ &  & 2.2       & >5        & 0.76      & 1.2       & >5        &  &  & >5       \\ 
\ettc $\:(f_{\rm gap} = 0.5)$ &  & 2         & >5        & 0.51      & 0.99      & >5        &  &  & >5       \\ 
\ettc $\:(f_{\rm gap} = 0.25)$ &  & 1.8       & >5        & 0.27      & 0.75      & >5        &  & 0.25      & >5       \\ 
\ettc $\:(f_{\rm gap} = 0.083)$ & 0.092     & 1.6       & >5        & 0.1       & 0.58      & >5        &  & 0.091     & >5       \\ 
\ettc  & 0.021     & 1.6       & >5        & 0.017     & 0.49      & >5        & 0.00096   & 0.012     & >5       \\ 
\hline
\hline
\etls  & 0.06      & 4.3       & >5        & 0.073     & 2.6       & >5        & 0.0015    & 0.017     & >5       \\ 
\etlp  & 0.061     & 4.3       & >5        & 0.17      & >5        & >5        & 0.0017    & 0.023     & >5       \\ 
\hline
\etl   & 0.062     & 4.3       & >5        & 1.1       & >5        & >5        & 0.0033    & 0.052     & >5       \\ 
\hline
\etlc $\:(f_{\rm gap} = 0.75)$ &  & 1.8       & >5        & 0.75      & 1.1       & >5        &  &  & >5       \\ 
\etlc $\:(f_{\rm gap} = 0.5)$ &  & 1.6       & >5        & 0.51      & 0.84      & >5        &  &  & >5       \\ 
\etlc $\:(f_{\rm gap} = 0.25)$ &  & 1.3       & >5        & 0.26      & 0.56      & >5        &  &  & >5       \\ 
\etlc $\:(f_{\rm gap} = 0.083)$ &  & 1.2       & >5        & 0.093     & 0.4       & >5        &  &  & 4.9      \\ 
\etlc  & 0.016     & 1.2       & >5        & 0.01      & 0.33      & >5        & 0.00065   & 0.0065    & 5        \\ 
\hline
\hline
\ces   & 0.033     & 2.3       & >5        & 1.5       & >5        & >5        & 0.057     & 2.7       & >5       \\ 
\cep   & 0.034     & 2.3       & >5        & 5         & >5        & >5        & 0.11      & >5        & >5       \\ 
\hline
\ce    & 0.034     & 2.3       & >5        & >5        & >5        & >5        & >5        & >5        & >5       \\ 
\hline
\cet $\:(f_{\rm gap} = 0.75)$ &  & 1.8       & >5        & 0.77      & 1.2       & >5        & 0.75      & 0.76      & >5       \\ 
\cet $\:(f_{\rm gap} = 0.5)$ &  & 1.7       & >5        & 0.52      & 0.99      & >5        & 0.5       & 0.51      & >5       \\ 
\cet $\:(f_{\rm gap} = 0.25)$ &  & 1.6       & >5        & 0.27      & 0.75      & >5        & 0.25      & 0.26      & >5       \\ 
\cet $\:(f_{\rm gap} = 0.083)$ &  & 1.6       & >5        & 0.1       & 0.58      & >5        & 0.084     & 0.094     & >5       \\ 
\cet   & 0.021     & 1.6       & >5        & 0.017     & 0.49      & >5        & 0.00096   & 0.012     & >5       \\ 
\hline
\cel $\:(f_{\rm gap} = 0.75)$ &  & 1.5       & >5        & 0.76      & 1.1       & >5        & 0.75      & 0.76      & >5       \\ 
\cel $\:(f_{\rm gap} = 0.5)$ &  & 1.4       & >5        & 0.51      & 0.84      & >5        & 0.5       & 0.51      & >5       \\ 
\cel $\:(f_{\rm gap} = 0.25)$ &  & 1.3       & >5        & 0.26      & 0.56      & >5        & 0.25      & 0.26      & >5       \\ 
\cel $\:(f_{\rm gap} = 0.083)$ &  & 1.2       & >5        & 0.094     & 0.4       & >5        & 0.084     & 0.089     & 4.9      \\ 
\cel   & 0.016     & 1.2       & >5        & 0.01      & 0.33      & >5        & 0.00065   & 0.0065    & 5        \\ 
\hline
\hline
\hline
 & \multicolumn{3}{c|}{\ewsnrth} & \multicolumn{3}{c|}{$\ewskyth \:\: [{\rm deg^2}]$} & \multicolumn{3}{c}{$\volth \:\: [{\rm Mpc^3}]$} \\ 
 & 20 & 80 & 160 & 10 & 1 & 0.1 & $1000$ & $100$ & $10$ \\ 
\hline
\hline
\etts  & 0.037     & 2.4       & >5        & >5        & >5        & >5        & >5        & >5        & >5       \\ 
\ettp  & 0.037     & 2.4       & >5        & >5        & >5        & >5        & >5        & >5        & >5       \\ 
\hline
\ett   & 0.037     & 2.4       & >5        & >5        & >5        & >5        & >5        & >5        & >5       \\ 
\hline
\ettc $\:(f_{\rm gap} = 0.75)$ &  & 1         & 3.2       & 1.8       & >5        & >5        & 4         & >5        & >5       \\ 
\ettc $\:(f_{\rm gap} = 0.5)$ &  & 0.77      & 3.3       & 1.5       & >5        & >5        & 3.6       & >5        & >5       \\ 
\ettc $\:(f_{\rm gap} = 0.25)$ &  & 0.54      & 2.8       & 1.3       & >5        & >5        & 3.3       & >5        & >5       \\ 
\ettc $\:(f_{\rm gap} = 0.083)$ &  & 0.4       & 2.7       & 1.2       & >5        & >5        & 2.9       & >5        & >5       \\ 
\ettc  & 0.0075    & 0.32      & 2.6       & 1.1       & >5        & >5        & 2.8       & >5        & >5       \\ 
\hline
\hline
\etls  & 0.019     & 1.2       & >5        & >5        & >5        & >5        & >5        & >5        & >5       \\ 
\etlp  & 0.019     & 1.2       & >5        & >5        & >5        & >5        & >5        & >5        & >5       \\ 
\hline
\etl   & 0.019     & 1.2       & >5        & >5        & >5        & >5        & >5        & >5        & >5       \\ 
\hline
\etlc $\:(f_{\rm gap} = 0.75)$ &  & 0.85      & 2.7       & 1.3       & >5        & >5        & 2.8       & >5        & >5       \\ 
\etlc $\:(f_{\rm gap} = 0.5)$ &  & 0.65      & 2.5       & 1.1       & >5        & >5        & 2.5       & >5        & >5       \\ 
\etlc $\:(f_{\rm gap} = 0.25)$ &  & 0.46      & 2.4       & 0.9       & >5        & >5        & 2.4       & >5        & >5       \\ 
\etlc $\:(f_{\rm gap} = 0.083)$ &  & 0.33      & 2.2       & 0.77      & >5        & >5        & 2.2       & >5        & >5       \\ 
\etlc  & 0.006     & 0.27      & 2.2       & 0.66      & >5        & >5        & 2.1       & >5        & >5       \\ 
\hline
\hline
\ces   & 0.011     & 0.48      & 3.7       & >5        & >5        & >5        & >5        & >5        & >5       \\ 
\cep   & 0.011     & 0.48      & 3.7       & >5        & >5        & >5        & >5        & >5        & >5       \\ 
\hline
\ce    & 0.011     & 0.48      & 3.7       & >5        & >5        & >5        & >5        & >5        & >5       \\ 
\hline
\cet $\:(f_{\rm gap} = 0.75)$ &  &  & 2.8       & 1.8       & >5        & >5        & 4         & >5        & >5       \\ 
\cet $\:(f_{\rm gap} = 0.5)$ &  &  & 2.8       & 1.6       & >5        & >5        & 3.6       & >5        & >5       \\ 
\cet $\:(f_{\rm gap} = 0.25)$ &  & 0.4       & 2.7       & 1.3       & >5        & >5        & 3.3       & >5        & >5       \\ 
\cet $\:(f_{\rm gap} = 0.083)$ &  & 0.36      & 2.6       & 1.2       & >5        & >5        & 2.9       & >5        & >5       \\ 
\cet   & 0.0075    & 0.32      & 2.6       & 1.1       & >5        & >5        & 2.8       & >5        & >5       \\ 
\hline
\cel $\:(f_{\rm gap} = 0.75)$ &  &  & 2.4       & 1.4       & >5        & >5        & 2.8       & >5        & >5       \\ 
\cel $\:(f_{\rm gap} = 0.5)$ &  &  & 2.4       & 1.2       & >5        & >5        & 2.5       & >5        & >5       \\ 
\cel $\:(f_{\rm gap} = 0.25)$ &  & 0.37      & 2.4       & 0.94      & >5        & >5        & 2.4       & >5        & >5       \\ 
\cel $\:(f_{\rm gap} = 0.083)$ &  & 0.31      & 2.2       & 0.77      & >5        & >5        & 2.2       & >5        & >5       \\ 
\cel   & 0.006     & 0.27      & 2.2       & 0.66      & >5        & >5        & 2.1       & >5        & >5       \\ 
\hline
\hline
\end{tabular}
\label{tab:timings_bns}
}
\end{center}
\end{table*}

\begin{table*}[]

\caption{Companion table for estimated observation times \tobs for BNSs collecting the same information as shown across Figs.~\ref{fig:timings_N_10} and \ref{fig:timings_N_10_india}. Each entry in the table corresponds to a marker in the figures for detector and network configurations \cet, \cetp, \cets, \cel, \celp, and \cels.}

\begin{center}
\resizebox{0.63\textwidth}{!}
{\centering
\begin{tabular}{l|ccc|ccc|ccc}
\hline 
\hline 
\multirow{2}{*}{$\bm{T_{\rm obs}\mbbh \:\:[{\rm yr}]}$} & \multicolumn{3}{c|}{\snrth} & \multicolumn{3}{c|}{$\skyth \:\: [{\rm deg^2}]$} & \multicolumn{3}{c}{\disth} \\ 
 & 100 & 300 & 1000 & 10 & 1 & 0.1 & 0.1 & 0.01 & 0.005 \\ 
\hline 
\hline 
\cet   & 0.002     & 0.048     & 2.1       & 0.00044   & 0.0077    & 0.21      & 0.0003    & 0.069     & 0.7      \\ 
\cetp  & 0.002     & 0.048     & 2.1       & 0.00033   & 0.0049    & 0.14      & 0.00028   & 0.062     & 0.66     \\ 
\cets  & 0.002     & 0.047     & 2.1       & 0.00028   & 0.0033    & 0.081     & 0.00027   & 0.054     & 0.63     \\ 
\hline 
\cel   & 0.0016    & 0.038     & 1.5       & 0.00033   & 0.005     & 0.13      & 0.00023   & 0.043     & 0.44     \\ 
\celp  & 0.0016    & 0.038     & 1.5       & 0.00028   & 0.0037    & 0.1       & 0.00022   & 0.039     & 0.4      \\ 
\cels  & 0.0016    & 0.038     & 1.5       & 0.00023   & 0.0026    & 0.062     & 0.00021   & 0.036     & 0.38     \\ 
\hline 
\hline 
\hline 
 & \multicolumn{3}{c|}{\pmsnrth} & \multicolumn{3}{c|}{} & \multicolumn{3}{c}{$\volth \:\: [{\rm Mpc^3}]$} \\ 
 & 20 & 60 & 200 &  &  &  & $100$ & $10$ & $1$ \\ 
\hline 
\hline 
\cet   & 0.0029    & 0.096     & >5        &  &  &  & 1.9       & >5        & >5       \\ 
\cetp  & 0.0029    & 0.096     & >5        &  &  &  & 1.5       & 4.3       & >5       \\ 
\cets  & 0.0028    & 0.093     & >5        &  &  &  & 1.2       & 3.3       & >5       \\ 
\hline 
\cel   & 0.0022    & 0.061     & >5        &  &  &  & 1.4       & 4.3       & >5       \\ 
\celp  & 0.0022    & 0.06      & >5        &  &  &  & 1.2       & 3.7       & >5       \\ 
\cels  & 0.0021    & 0.057     & >5        &  &  &  & 1         & 2.8       & >5       \\ 
\hline 
\hline 
\hline 
\hline 
\multirow{2}{*}{$\bm{T_{\rm obs}\mbns \:\:[{\rm yr}]}$} & \multicolumn{3}{c|}{\snrth} & \multicolumn{3}{c|}{$\skyth \:\: [{\rm deg^2}]$} & \multicolumn{3}{c}{\disth} \\ 
 & 50 & 200 & 400 & 10 & 1 & 0.1 & 0.2 & 0.1 & 0.01 \\ 
\hline 
\hline 
\cet   & 0.021     & 1.6       & >5        & 0.017     & 0.49      & >5        & 0.00096   & 0.012     & >5       \\ 
\cetp  & 0.021     & 1.6       & >5        & 0.0087    & 0.27      & >5        & 0.00086   & 0.0091    & >5       \\ 
\cets  & 0.021     & 1.4       & >5        & 0.0048    & 0.14      & 5         & 0.00081   & 0.008     & >5       \\ 
\hline 
\cel   & 0.016     & 1.2       & >5        & 0.01      & 0.33      & >5        & 0.00065   & 0.0065    & 5        \\ 
\celp  & 0.016     & 1.2       & >5        & 0.0063    & 0.19      & >5        & 0.00062   & 0.0054    & 4.3      \\ 
\cels  & 0.016     & 1.2       & >5        & 0.004     & 0.12      & 3.7       & 0.00058   & 0.0049    & 4.3      \\ 
\hline 
\hline 
\hline 
 & \multicolumn{3}{c|}{\ewsnrth} & \multicolumn{3}{c|}{$\ewskyth \:\: [{\rm deg^2}]$} & \multicolumn{3}{c}{$\volth \:\: [{\rm Mpc^3}]$} \\ 
 & 20 & 80 & 160 & 10 & 1 & 0.1 & $1000$ & $100$ & $10$ \\ 
\hline 
\hline 
\cet   & 0.0075    & 0.32      & 2.6       & 1.1       & >5        & >5        & 2.8       & >5        & >5       \\ 
\cetp  & 0.0075    & 0.32      & 2.6       & 1         & >5        & >5        & 2.1       & >5        & >5       \\ 
\cets  & 0.0075    & 0.32      & 2.6       & 1         & >5        & >5        & 1.7       & >5        & >5       \\ 
\hline 
\cel   & 0.006     & 0.27      & 2.2       & 0.66      & >5        & >5        & 2.1       & >5        & >5       \\ 
\celp  & 0.006     & 0.27      & 2.2       & 0.66      & >5        & >5        & 1.6       & >5        & >5       \\ 
\cels  & 0.006     & 0.27      & 2.2       & 0.63      & >5        & >5        & 1.4       & >5        & >5       \\ 
\hline 
\hline 
\end{tabular} 
\label{tab:india_bns}
}
\end{center}
\end{table*}

\section{Full observation time maps} \label{app:maps}

\begin{figure*}[p]
    \centering
    \includegraphics[width=\linewidth]{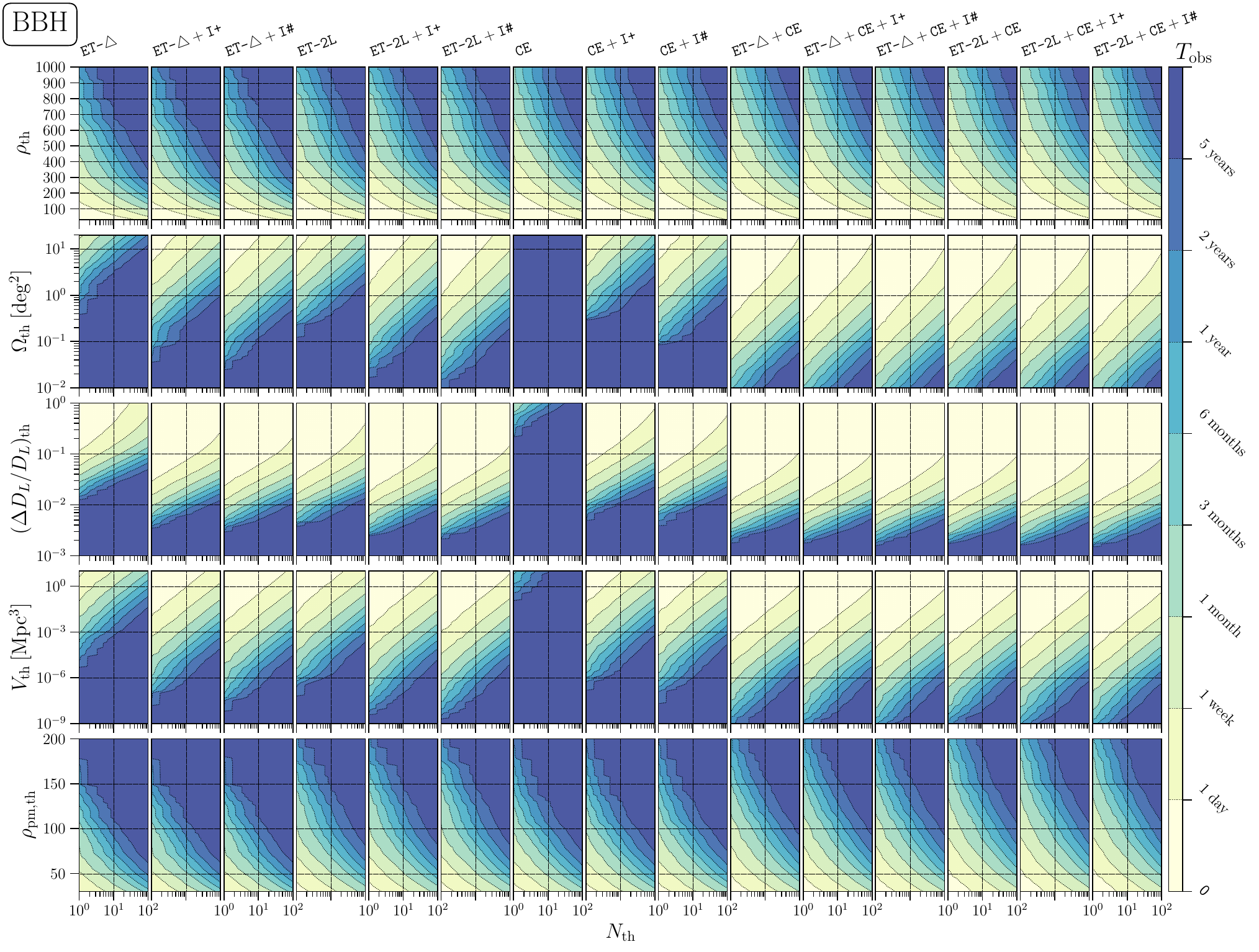}

    \caption{Maps of expected observation times \tobs to reach a target of at
    least \nth BBH signals satisfying a threshold for SNR \snrth,
    post-merger SNR \pmsnrth, sky area \skyth, relative luminosity distance
    error \disth, or comoving error volume \volth, observed with \num{15}
    different XG, ground-based GW detector networks. Quantities \nth, \snrth, \pmsnrth\
    and \skyth, \disth, \volth represent lower and upper thresholds,
    respectively. The observing time \tobs is calculated for a fiducial cosmic merger rate of
    $\mathcal{R}=10^5\,{\rm yr^{-1}}$ and is binned into \num{eight}
    human-experiment-scale time frames, with the \num{nineth} bin corresponding
    to observation times exceeding \num{five} years. The merger events are
    Poisson-distributed and 100-fold bootstrapped from $10^6$ simulated
    signals.}

    \label{fig:bbh}
\end{figure*}

\begin{figure*}[p]
    \centering
    \includegraphics[width=\linewidth]{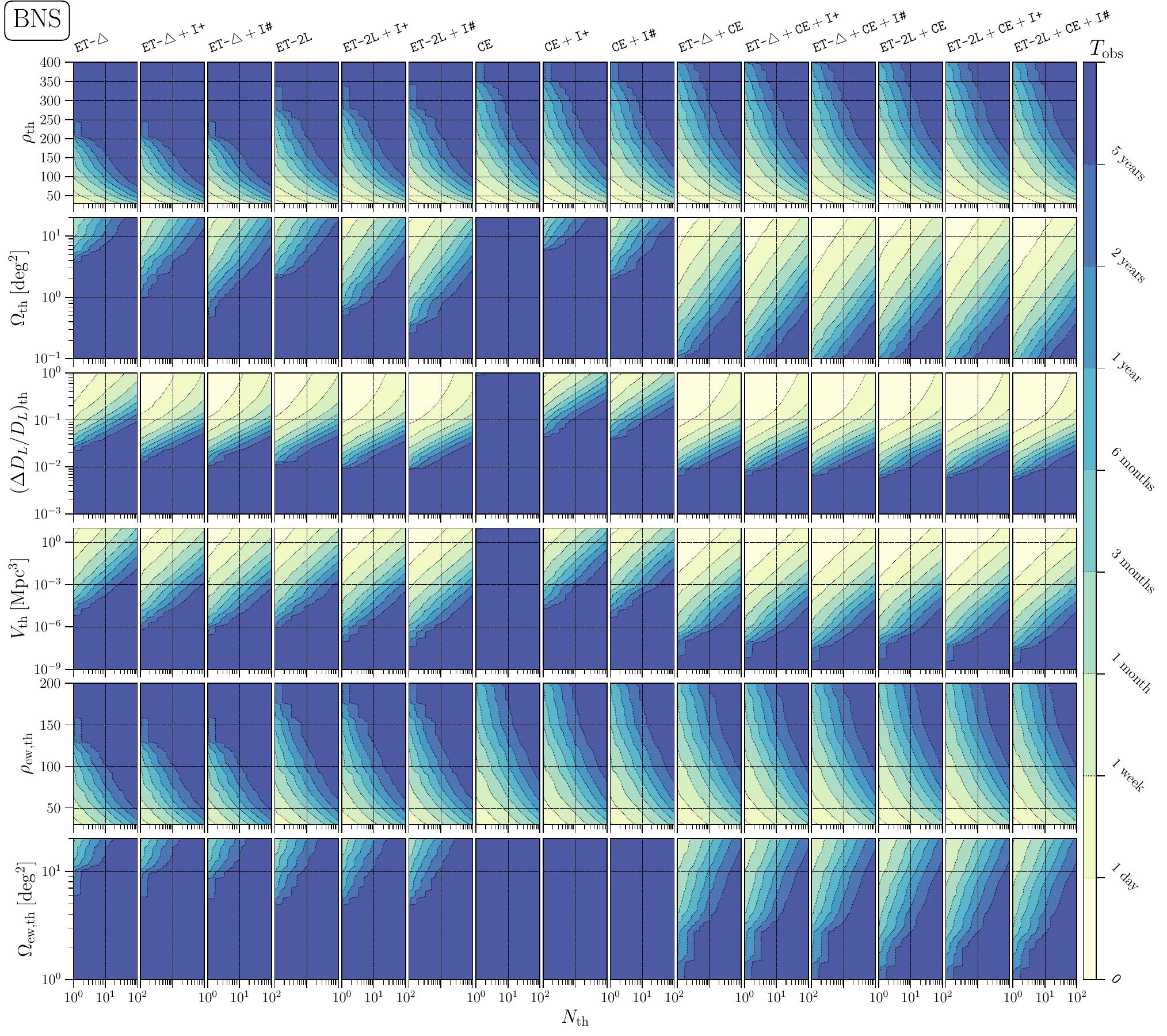}

    \caption{Maps of expected observation times \tobs to reach a target of at
    least \nth BBH signals satisfying a threshold for SNR \snrth,
    early-warning SNR \ewsnrth, sky area \skyth, early-warning sky area
    \ewskyth, relative luminosity distance error \disth, or comoving error
    volume \volth, observed with \num{15} different XG, ground-based GW detector
    networks. Quantities \nth, \snrth, \ewsnrth and \skyth, \ewskyth, \disth, \volth\
    represent lower and upper thresholds, respectively. The observing time \tobs is calculated for
    a fiducial cosmic merger rate of $\mathcal{R}=10^5\,{\rm yr^{-1}}$ and is
    binned into \num{eight} human-experiment-scale time frames, with the
    \num{nineth} bin corresponding to observation times exceeding \num{five}
    years. The merger events are Poisson-distributed and 100-fold bootstrapped
    from $10^6$ simulated signals.}

    \label{fig:bns}
\end{figure*}

\begin{figure*}[p]
    \centering
    \includegraphics[width=\linewidth]{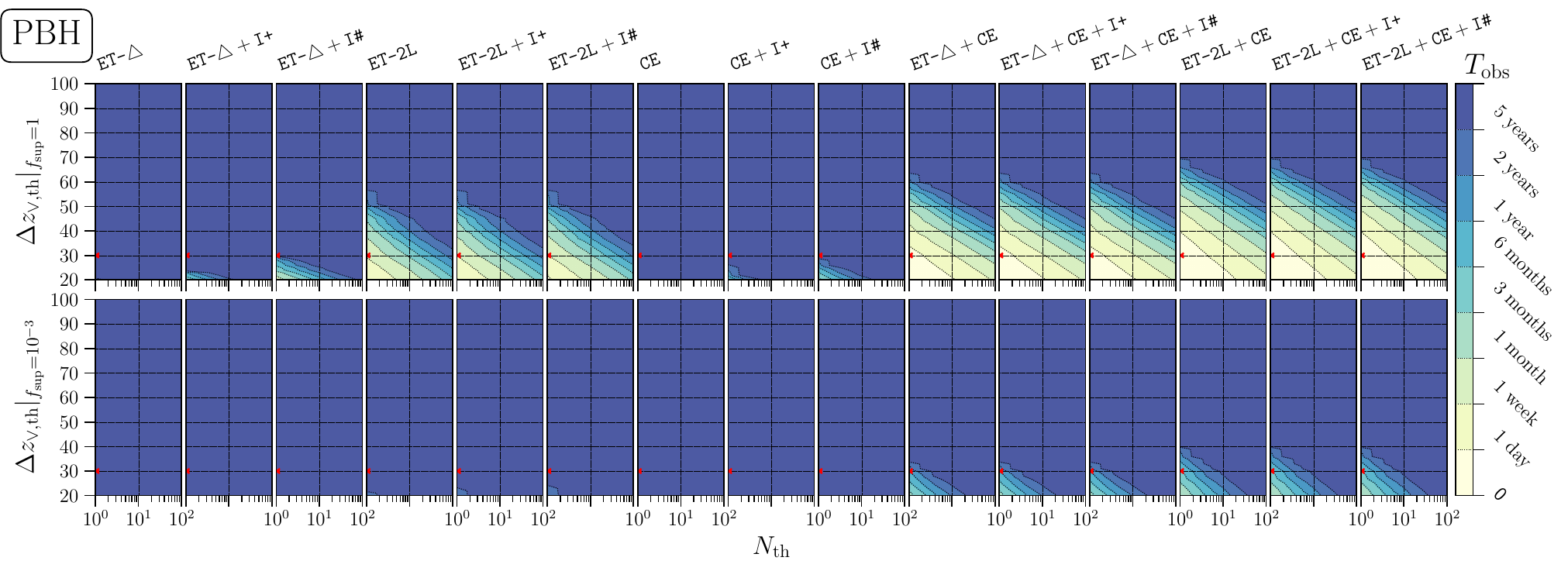}

    \caption{Maps of expected observation times \tobs to reach a
    target of at least \nth PBH signals satisfying varying thresholds for the
    lower redshift error bound \zloth at two values for the PBH-formation
    suppression factor $f_{\rm sup}\in\{10^{-3},1\}$, observed with \num{15}
    different XG, ground-based GW detector networks.
    The observing time \tobs is calculated for a fiducial cosmic merger rate of
    $\mathcal{R}=2\times10^5\,{\rm yr^{-1}}$ and is binned into \num{eight}
    human-experiment-scale time frames, with the \num{nineth} bin corresponding
    to observation times exceeding \num{five} years. The merger events are
    Poisson-distributed and 100-fold bootstrapped from $10^6$ simulated signals.
    The red marker indicates the corresponding positions, $\nth=1$ and
    $\zloth=30$ corresponding to Fig.~\ref{fig:timings_N_1_pbh}.} 

    \label{fig:pbh}
\end{figure*}

Figures~\ref{fig:bbh}, \ref{fig:bns}, and \ref{fig:pbh} present full maps of
\nth against a range of threshold values for all the studied metrics for BBH,
BNS, and PBH mergers: the SNR \snrth, sky area \skyth, relative luminosity
distance error \disth, comoving error volume \volth, post-merger SNR \pmsnrth
for BBHs, early-warning SNR \ewsnrth and early-warning sky area \ewskyth for
BNSs, and the lower redshift error bound \zloth for PBHs. The maps are binned
into \num{eight} human-experiment-scale time frames, with the \num{nineth} bin
corresponding to observation times exceeding \num{five} years.

\clearpage

\section*{References}
\bibliographystyle{JHEP}
\bibliography{3g_timing}

\end{document}